\documentclass[
    reprint,
    amsmath,
    amssymb,
    aps,
    pra,
    superscriptaddress,
    floatfix
]{revtex4-2}

\usepackage{tikz}
\usepackage{graphicx}
\usepackage{dcolumn}
\usepackage{bm}
\usetikzlibrary{shapes.geometric}

\begin{document}

\title{Beyond Feasibility: Finite-Depth Accessibility in Constrained QAOA}

\author{Rushikesh Ubale}
\email{rushikeshubale.work@gmail.com}
\affiliation{Qkrishi Quantum, India}

\author{Gregory T. Byrd}
\email{gbyrd@ncsu.edu}
\affiliation{North Carolina State University, Raleigh, North Carolina, USA}

\author{Yasar Mulani}
\email{mulaniyhofficial@gmail.com}
\affiliation{Qkrishi Quantum, India}

\author{Sangram Deshpande}
\email{ssdesh24@ncsu.edu}
\affiliation{North Carolina State University, Raleigh, North Carolina, USA}


\begin{abstract}

Constraint-preserving mixers allow the Quantum Approximate Optimization
Algorithm (QAOA) to remain within a feasible subspace, but feasibility and
global mixer connectivity do not determine which feasible configurations are
available to a particular shallow circuit. We study constrained QAOA at the
level of the circuit actually executed: a specified feasible initial state,
finite depth, and finite schedule of mixer interactions. We formulate the
finite-depth accessible set $R_p(M,x_0)$ and show that the computational-basis
support of the corresponding ideal QAOA circuit is contained within this set.
This yields a parameter-independent objective ceiling
$f_R^\star=\max_{x\in R_p}f(x)$ and a normalized reachable-quality diagnostic
$Q_R$, separating the amount of feasible space exposed by a schedule from the
objective quality contained within it.

We evaluate this distinction using resource-matched block-local XY mixer
schedules for constrained portfolio optimization and synthetic
block-constrained quadratic problems. At fixed depth and comparable executed
mixer-interaction budgets, different schedules can expose accessible sets of
similar or greater size while containing substantially different objective
quality. Across 207 favorable-tail CVaR QAOA configurations, $Q_R$ shows the
strongest association with realized shallow-QAOA performance among the
structural diagnostics considered, with Pearson and Spearman correlations of
$0.7003$ and $0.8057$, respectively; the relationship remains positive under
controlled and dependence-aware analyses. The distinction persists under
changes in feasible initialization, portfolio size, and objective family.

A compilation-only study further shows that, for the dense portfolio
Hamiltonians considered here, phase-separator routing dominates the compiled
two-qubit resource scale, while mixer schedules with comparable physical cost
can still exhibit markedly different finite-depth accessible quality. These
results motivate objective-relevant finite-depth accessibility as a
complementary diagnostic for designing and evaluating resource-bounded
constrained-QAOA circuits.

\end{abstract}

\maketitle

\section{Introduction}
\label{sec:introduction}

The Quantum Approximate Optimization Algorithm (QAOA) and its generalization
to the Quantum Alternating Operator Ansatz provide a flexible framework for
encoding combinatorial optimization problems into parameterized quantum
circuits~\cite{Farhi2014,hadfield2019alternating}. For unconstrained problems,
the standard transverse-field mixer allows amplitude to propagate through the
computational basis without regard to feasibility. Many practical optimization
problems, however, impose hard combinatorial constraints such as fixed
cardinality, assignment, scheduling, partitioning, or resource-allocation
requirements. Allowing a circuit to explore infeasible configurations can
waste sampling probability and can require penalty terms whose scales must be
balanced against the objective. Constraint-preserving mixers address this by
restricting the dynamics to a feasible subspace and embedding part of the
problem structure directly into the ansatz
~\cite{hadfield2019alternating,wang2020xy,fuchs2022constrained}.

This work focuses on a more specific finite-depth question. Suppose a
constrained QAOA circuit starts from a feasible computational-basis state
$x_0$, has depth $p$, and executes a particular sequence of mixer rounds
$M=(M_1,\ldots,M_p)$. Even when the available mixer interactions connect the
full feasible manifold in principle, a shallow circuit uses only a finite
subset of those interactions. This leads to the question: \emph{which feasible basis
configurations can this actual circuit structurally access from $x_0$ at the
available depth?} We call this \emph{finite-depth accessibility} and denote
the corresponding set by $R_p(M,x_0)\subseteq\mathcal F$.

For the block-local disjoint XY mixers studied here, a single round can contain
several disjoint exchanges. The accessible-set construction therefore
propagates all compatible subsets of the active exchanges in each executed
round. It is not a one-edge-at-a-time walk on a static transition graph. This
construction gives a parameter-independent restriction on the circuit: for
the class of circuits considered here,
\begin{equation}
    \operatorname{supp}(P_{\boldsymbol{\beta},\boldsymbol{\gamma}})
    \subseteq R_p(M,x_0).
    \label{eq:intro_support}
\end{equation}
A feasible state outside $R_p$ has zero sampling probability for every choice
of variational parameters under the fixed schedule. If the global optimum is
outside this set, changing the classical optimizer or adding measurement shots
cannot make that state available without changing the circuit structure.

For optimization, the size of $R_p$ is not enough. Two schedules can expose
similar numbers of feasible states while making very different objective
values available. We define the best structurally accessible objective and its
normalized quality as
\begin{equation*}
    f_R^\star=\max_{x\in R_p(M,x_0)}f(x),
    \qquad
    Q_R=\frac{f_R^\star-f(x_0)}{f^\star-f(x_0)}.
\end{equation*}
Using $f(x_0)$ as the lower reference is deliberate: the accessible set itself
is defined relative to the chosen starting state, so $Q_R$ measures the
fraction of the objective gap from that starting point to the global feasible
optimum that is structurally available. With this convention, $Q_R=0$ means
that the schedule exposes no improvement over $x_0$, while $Q_R=1$ means that
a globally optimal feasible objective is present in the accessible set.
$Q_R$ does not specify the probability of sampling a good solution and does
not describe interference, trainability, or optimizer behavior. It measures
the quality of the opportunity created by the finite-depth schedule.

\begin{figure}[t]
    \centering
    \includegraphics[width=0.95\columnwidth]{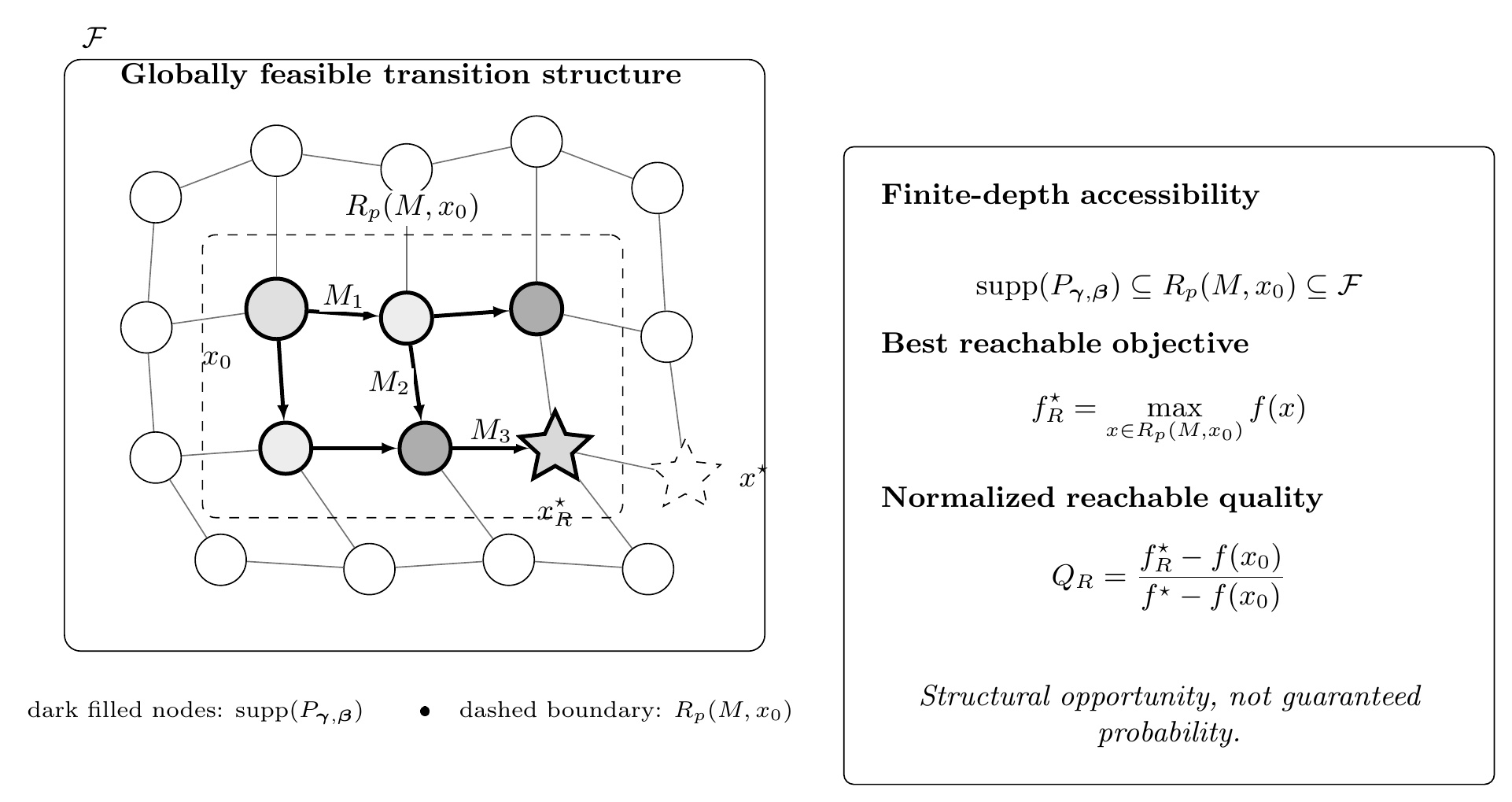}
    \caption{Conceptual distinction between global feasible connectivity and
    finite-depth accessibility. Although the full constraint-preserving mixer
    structure may connect the feasible manifold $\mathcal{F}$, a circuit
    initialized at $x_0$ and executing only a finite mixer schedule can
    structurally access only $R_p(M,x_0)\subseteq\mathcal{F}$. The accessible
    set contains the computational-basis support of the ideal circuit and
    defines the best reachable objective $f_R^\star$ and normalized reachable
    quality $Q_R$.}
    \label{fig:finite_depth_accessibility}
\end{figure}

The question sits within an established literature on constrained mixer
design. XY constructions provide natural dynamics for fixed-Hamming-weight and
related feasible spaces~\cite{wang2020xy}, and more general constructions
characterize mixers that preserve arbitrary feasible subspaces and the
transitions they induce~\cite{fuchs2022constrained}. Prior studies have also
examined resource-efficient implementations, initialization, and mixer
topology. The interaction between the initial state and a constrained mixer
can materially affect shallow-QAOA performance~\cite{He2023}, and recent
Lie-algebraic analyses show that different XY-mixer topologies can generate
qualitatively different dynamical structures~\cite{kordonowy2026lie}.
Selective and adaptive allocation of mixer operations has also been explored
as a way to reduce circuit resources while retaining useful optimization
behavior~\cite{ni2026adaptive}.

Finite circuit depth introduces another set of limitations. Shallow QAOA is
subject to locality- and symmetry-induced restrictions, and ``reachability''
has been used in several contexts to describe limits on what a variational
ansatz can represent or optimize~\cite{akshay2020reachability,wU2026reachability}.
Recent work on constrained optimization has studied finite-depth success
conditions and separated mixer-induced amplitude transport from coherent
interference and post-processing effects~\cite{onah2026finite,onah2026geometry}.
Transition graphs have likewise been used to characterize the feasible
transitions generated by constrained mixers~\cite{gupta2026rostering}. 
A recent analysis of constrained QAOA approaches reachability from a complementary expressivity and quantum-control perspective. Using dynamical Lie algebras and feasible-subspace state orbits, it shows that two mixers with the same feasible-state transitions can still have different dynamical expressivity \cite{tsvelikhovskiy2026expressivepowerconstrainedqaoa}.

Our focus is different. We condition on a particular feasible
computational-basis initialization, a fixed shallow depth, and the actual
finite sequence of mixer interactions that is executed. We then ask which
computational-basis configurations are structurally accessible under that
schedule and how much objective quality is contained in the resulting
finite-depth accessible set.

The main contributions are as follows.
\begin{enumerate}
    \item \textbf{Finite-depth accessibility.}
    We formulate schedule- and initial-state-specific accessibility for
    block-constrained QAOA, derive the accessible set generated by executed
    disjoint XY rounds, and establish support containment. States outside this
    set cannot be sampled by the specified ideal circuit for any choice of
    variational parameters.

    \item \textbf{Objective quality inside the accessible set.}
    We use $f_R^\star$ and $Q_R$ to distinguish accessible volume from the
    objective quality contained within that volume. A larger accessible set or
    broader transition coverage does not, by itself, imply a higher reachable
    objective ceiling.

    \item \textbf{Resource-controlled empirical analysis.}
    We compare several constrained mixer schedules at matched executed
    XY-interaction budgets and relate structural accessibility metrics to
    realized QAOA performance. The analysis is repeated across feasible
    initial states, nested portfolio sizes, and synthetic block-QUBO instances,
    with dependence-aware statistics used to account for repeated observations
    from the same underlying problems.

    \item \textbf{Application and compilation analysis.}
    We evaluate the framework on a market-grounded portfolio benchmark with
    sector constraints, shrinkage-based financial estimation, and turnover
    cost, and we complement the algorithmic study with topology-constrained
    compilation. For the dense portfolio Hamiltonians considered here, the
    phase-separation layer sets the dominant compiled two-qubit resource scale,
    even though different mixer schedules can expose very different objective
    quality.
\end{enumerate}

To test these ideas, we use fixed, rotating-disjoint, random-disjoint, and
objective-improving mixer schedules while controlling the number of executed
XY-pair interactions. The objective-improving construction is not presented
as a universally superior or adaptive mixer; selective mixer allocation has
been studied previously~\cite{ni2026adaptive}. Here it serves as a controlled
way to change the objective content of the accessible region without giving
one schedule a larger interaction budget.

The primary application is a stylized portfolio-selection benchmark built
from approximately a decade of historical U.S. equity prices. Portfolio
optimization commonly balances expected return, risk, allocation constraints,
and implementation costs~\cite{Markowitz1952,CornuejolsTutuncu2007}. Our
executable pipeline operates on 120 loaded equity price series and constructs
a 20-asset instance with five sector blocks, four candidate assets per block,
and exactly two selected assets in each block. Expected returns are shrunk
toward the cross-sectional mean, risk is estimated with a Ledoit--Wolf
covariance estimator, and the objective includes a turnover penalty relative
to a fixed reference portfolio. The resulting feasible space contains
$\prod_{g=1}^{5}\binom{4}{2}=7776$ configurations. We also repeat the analysis
for different feasible initial states, nested portfolio sizes, and synthetic
block-constrained quadratic objectives so that the accessibility result is not
assessed only on one financial instance.

Across these experiments, structural coverage and objective-relevant
accessibility behave differently. Schedules that expose different fractions
of the feasible space can have comparable objective ceilings, while relatively
small accessible sets can contain high-quality solutions. Under the tested
resource-matched settings, $Q_R$ has the strongest association with realized
shallow-QAOA performance among the structural diagnostics considered. We do
not treat this association as a universal performance law. It supports the
more limited conclusion that feasibility, connectivity, and reachable volume
can miss an important feature of a shallow constrained circuit: the objective
quality of the region its executed schedule actually makes available.

The resulting progression is straightforward. Feasibility asks whether the
circuit remains inside $\mathcal F$; global connectivity asks what the mixer
can connect in principle; finite-depth accessibility asks what the executed
schedule can expose from $x_0$; and objective-relevant accessibility asks
whether that exposed region contains useful solutions. Variational
optimization and sampling act within this structural opportunity, but they
cannot recover objective quality that the fixed circuit never makes
accessible.
\section{Finite-Depth Accessibility in Constrained QAOA}
\label{sec:accessibility}

Constraint-preserving mixers are usually specified by the feasible
subspace they preserve and by the transitions they permit within that
subspace. For a finite-depth circuit, however, the mixer Hamiltonian alone
does not determine the states available to the algorithm. The initial state,
the particular interactions retained in each mixer round, and the order in
which those rounds are executed also matter.

This section makes that dependence explicit. We work at the level of
computational-basis configurations and define the feasible states that can be
accessed by the \emph{executed} mixer schedule. The resulting construction is
combinatorial and independent of the variational angles. This gives a structural restriction of the circuit before parameter
optimization or sampling is considered.

\subsection{Feasible States and Executed Mixer Schedules}
\label{subsec:feasible_schedule}

Let
\begin{equation*}
    \mathcal{F}\subseteq\{0,1\}^{n}
\end{equation*}
denote the feasible computational-basis configurations of a constrained
optimization problem, and let
\begin{equation*}
    x_0\in\mathcal{F}
\end{equation*}
be the basis state used to initialize the QAOA circuit.

We consider a depth-$p$ circuit of the form
\begin{equation}
    |\psi_p(\boldsymbol{\beta},\boldsymbol{\gamma})\rangle
    =
    \prod_{\ell=1}^{p}
    U_M^{(\ell)}(\beta_\ell)
    U_C(\gamma_\ell)
    |x_0\rangle,
    \label{eq:qaoa_state}
\end{equation}
where $U_C(\gamma_\ell)$ is diagonal in the computational basis and
$U_M^{(\ell)}(\beta_\ell)$ is the constraint-preserving mixer executed at
layer $\ell$. Here $\boldsymbol{\gamma}=(\gamma_1,\ldots,\gamma_p)$ are
the cost-phase angles and $\boldsymbol{\beta}=(\beta_1,\ldots,\beta_p)$
are the mixer angles, following the standard QAOA parameterization
\cite{Farhi2014}. The mixer may vary from one layer to the next.

For the block-cardinality problems studied here, each mixer round consists of
a set of pairwise disjoint XY interactions,
\begin{equation*}
    M_\ell
    =
    \left\{(i,j)\right\}_{\ell},
\end{equation*}
with every pair contained inside a common constraint block. A pair
$(i,j)\in M_\ell$ is implemented through the standard XY exchange used in
constraint-preserving alternating-operator constructions
\cite{hadfield2019alternating}, with generator
\begin{equation}
    H^{XY}_{ij}
    =
    X_iX_j + Y_iY_j.
    \label{eq:xy_generator}
\end{equation}
Because XY evolution conserves Hamming weight on the interacting pair,
block-local applications preserve the corresponding block-cardinality
constraints.

The schedule
\begin{equation}
    M=(M_1,\ldots,M_p)
    \label{eq:mixer_schedule}
\end{equation}
is part of the circuit specification. Two circuits may therefore use the same
family of feasible XY interactions while exposing different finite-depth
transition structures.

For reference, the principal notation used below is summarized in
Table~\ref{tab:notation}.\\

\begingroup
\small
\refstepcounter{table}
\label{tab:notation}

\medskip
\noindent
\textbf{TABLE \thetable.} Principal notation for finite-depth accessibility.

\smallskip
\begin{center}
\begin{tabular}{ll}
\hline\hline
Symbol & Meaning \\
\hline
$n$ & number of binary decision variables \\
$\mathcal{F}$ & complete feasible set \\
$x_0$ & feasible initial basis configuration \\
$p$ & QAOA depth \\
$M_\ell$ & mixer round at layer $\ell$ \\
$M$ & complete mixer schedule \\
$A_\ell(x)$ & active exchange pairs at layer $\ell$ \\
$T_\ell(x)$ & configurations accessible in one round \\
$R_\ell$ & configurations accessible after $\ell$ rounds \\
$B$ & executed XY-pair interaction budget \\
$f(x)$ & classical maximization objective \\
$f^\star$ & globally optimal feasible objective \\
$f_R^\star$ & best objective value in $R_p$ \\
$Q_R$ & normalized best-reachable quality \\
$f_{\rm ref}$ & reporting reference used in $S_{\rm eff}$ \\
$S_{\rm eff}$ & realized normalized QAOA score \\
\hline\hline
\end{tabular}
\end{center}
\medskip
\endgroup

\subsection{Accessibility Under a Disjoint XY Round}
\label{subsec:round_accessibility}

A subtlety arises when a mixer layer contains several disjoint XY
interactions. Such a layer should not be interpreted as choosing one edge of
a transition graph. All disjoint interactions are present in the same unitary,
and each active exchange can contribute independently to the basis-state
support.

For a configuration $x$, define the active pairs in round $\ell$ as
\begin{equation}
    A_\ell(x)
    =
    \left\{
        (i,j)\in M_\ell :
        x_i\neq x_j
    \right\}.
    \label{eq:active_pairs}
\end{equation}
Pairs for which $x_i=x_j$ do not create a distinct exchanged configuration.

Since the pairs in $M_\ell$ are disjoint, the exchanges associated with any
subset
\begin{equation*}
    S\subseteq A_\ell(x)
\end{equation*}
can occur simultaneously without conflict. Let
\begin{equation*}
    \tau_S(x)
\end{equation*}
denote the bit string obtained by swapping the occupations on every pair in
$S$. The configurations structurally accessible from $x$ after that mixer
round are then
\begin{equation}
    T_\ell(x)
    =
    \left\{
        \tau_S(x):
        S\subseteq A_\ell(x)
    \right\}.
    \label{eq:round_transition}
\end{equation}
The empty subset is included, so $x\in T_\ell(x)$.

The finite-depth accessible sets are obtained recursively:
\begin{equation}
    R_0=\{x_0\},
    \label{eq:R0}
\end{equation}
and
\begin{equation}
    R_\ell
    =
    \bigcup_{x\in R_{\ell-1}}
    T_\ell(x),
    \qquad
    \ell=1,\ldots,p.
    \label{eq:R_recursion}
\end{equation}
We write $R_p(M,x_0)$ when the dependence on the complete schedule and initial
state needs to be shown explicitly.

\begin{figure*}[t]
    \centering
    \includegraphics[width=0.92\textwidth]{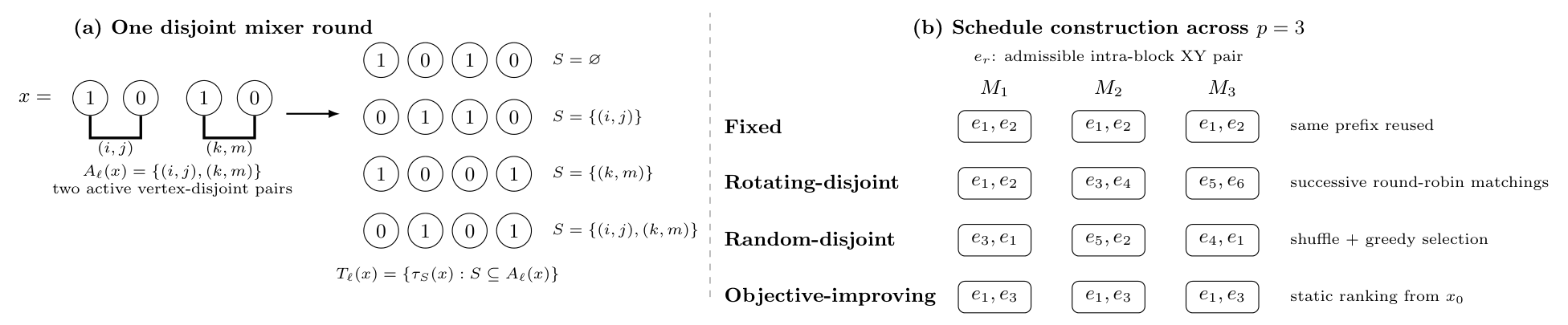}
    \caption{Disjoint-round semantics and mixer-schedule construction.
    (a) A single round can contain several vertex-disjoint XY exchanges, so all
    compatible subsets of the active exchanges contribute to the structural
    accessible set. (b) The fixed, rotating-disjoint, random-disjoint, and
    objective-improving schedule constructions used in the experiments; their
    precise definitions are given in Sec.~\ref{subsec:mixer_policies}.}
    \label{fig:mixer_semantics}
\end{figure*}

This construction differs from ordinary graph-distance reachability. A
single executed mixer round may generate several simultaneous exchanges and
can therefore produce configurations that would require multiple
single-edge moves on a static transition graph. Conversely, only the
interactions actually appearing in $M_1,\ldots,M_p$ contribute to
$R_p$. Connectivity of the full mixer graph consequently gives no guarantee
that the full feasible set is accessible at the depth being executed.

For $r=|A_\ell(x)|$ active disjoint pairs, a generic XY rotation has nonzero
amplitude on all $2^r$ subset-exchange outcomes provided neither the
sine nor cosine branch of the corresponding rotation vanishes. Thus a
single mixer layer can expand support exponentially in the number of
simultaneously active disjoint exchanges, although collisions between paths
and later schedule structure can reduce the number of distinct states.

\subsection{Support Containment}
\label{subsec:support_containment}

The accessible set provides a parameter-independent container for the
computational-basis support of the ideal circuit.

\paragraph{Proposition 1 (Feasibility preservation).}
If $x_0\in\mathcal{F}$ and every pair in every mixer round acts within a
constraint block whose Hamming weight must remain fixed, then
\begin{equation}
    R_\ell\subseteq\mathcal{F},
    \qquad
    \ell=0,\ldots,p.
    \label{eq:feasibility_prop}
\end{equation}

\emph{Proof.}
The claim holds trivially for $R_0$. An XY exchange permutes occupations
between two positions and therefore leaves the Hamming weight of their block
unchanged. Each element of $T_\ell(x)$ is obtained by applying a subset of
such disjoint exchanges to a feasible configuration $x$. It remains feasible,
and induction over $\ell$ proves the result.
\hfill$\square$

The diagonal cost unitary does not alter computational-basis support. The
mixer layer can only transfer amplitude among the subset-exchange outcomes
defined by Eq.~\eqref{eq:round_transition}. This gives the central structural
statement.

\paragraph{Proposition 2 (Finite-depth support containment).}
For the circuit in Eq.~\eqref{eq:qaoa_state},
\begin{equation}
    \operatorname{supp}
    \left(
        P_{\boldsymbol{\beta},\boldsymbol{\gamma}}
    \right)
    \subseteq
    R_p(M,x_0)
    \label{eq:support_containment}
\end{equation}
for every choice of variational parameters
$\boldsymbol{\beta},\boldsymbol{\gamma}$.

\emph{Proof.}
At depth zero the circuit has support only on $x_0$. Because $U_C$ is
diagonal, application of the cost unitary can change phases but cannot add a
new computational-basis state to the support. At mixer layer $\ell$, each
basis component $x$ can generate support only on configurations in
$T_\ell(x)$. Therefore the support after layer $\ell$ is contained in
\begin{equation*}
    \bigcup_{x\in R_{\ell-1}}T_\ell(x)=R_\ell.
\end{equation*}
Induction over the $p$ layers proves
Eq.~\eqref{eq:support_containment}.
\hfill$\square$

The containment can be strict. Particular mixer angles can suppress one or more
exchange branches, and amplitudes associated with different paths can
interfere destructively. We treat $R_p$ as a structural opportunity
set: states outside it are impossible under the specified schedule, while
states inside it are not guaranteed to appear with nonzero final probability.

An immediate consequence is that an inaccessible feasible solution cannot be
recovered by variational parameter optimization alone.

\paragraph{Corollary 1 (Inaccessible optimum).}
If a globally optimal feasible configuration $x^\star$ satisfies
\begin{equation*}
    x^\star\notin R_p(M,x_0),
\end{equation*}
then
\begin{equation*}
    P_{\boldsymbol{\beta},\boldsymbol{\gamma}}
    (x^\star)=0
\end{equation*}
for all values of the variational parameters.

The limitation is structural. Changing the classical optimizer, increasing
the number of optimizer iterations, or increasing the number of measurement
shots cannot make $x^\star$ available without modifying the initial state,
the mixer schedule, or the circuit depth.

\subsection{Reachable Objective Quality}
\label{subsec:reachable_quality}

Support containment gives a binary statement about whether a feasible
configuration is available. For optimization, a more useful question is how
good the best available configuration can be.

Let
\begin{equation}
    f^\star
    =
    \max_{x\in\mathcal{F}} f(x)
    \label{eq:global_best}
\end{equation}
denote the globally optimal feasible objective value. For a fixed
$(M,x_0,p)$, define
\begin{equation}
    f_R^\star
    =
    \max_{x\in R_p(M,x_0)} f(x).
    \label{eq:fRstar}
\end{equation}

Because every sample produced by the ideal circuit belongs to $R_p$,
\begin{equation*}
    f(X)\leq f_R^\star
\end{equation*}
for every computational-basis outcome $X$ with nonzero sampling
probability. Consequently,
\begin{equation}
    \mathbb{E}_{\boldsymbol{\beta},\boldsymbol{\gamma}}
    [f(X)]
    \leq
    f_R^\star.
    \label{eq:expected_bound}
\end{equation}
The same ceiling applies to the best sampled objective obtainable from the
specified ideal circuit.

For comparisons across initial states and problem instances, we use the
normalized quantity
\begin{equation}
    Q_R
    =
    \frac{
        f_R^\star-f(x_0)
    }{
        f^\star-f(x_0)
    },
    \label{eq:QR}
\end{equation}
whenever $f^\star\neq f(x_0)$. Under the maximization convention used here,
$Q_R=1$ means that the globally optimal feasible objective is contained in
the accessible set, while $Q_R=0$ means that the schedule exposes no
configuration better than the initial state.

Equation~\eqref{eq:QR} is intentionally structural. It contains no variational
parameters and does not describe how probability is distributed within
$R_p$. A schedule with $Q_R=1$ may still perform poorly because the
optimization landscape is difficult, favorable states receive little
amplitude, destructive interference suppresses them, or finite-shot sampling
fails to observe them. Conversely, a circuit cannot realize an objective
value above $f_R^\star$ without changing its structural specification.

Three questions are useful to keep separate:
\begin{equation*}
\begin{aligned}
\text{accessible fraction:}\quad
&\frac{|R_p|}{|\mathcal{F}|},
\\
\text{improving-state count:}\quad
&\left|\left\{x\in R_p:f(x)>f(x_0)\right\}\right|,
\\
\text{best accessible quality:}\quad
&Q_R.
\end{aligned}
\end{equation*}
These quantities can move in different directions.

\subsection{Connectivity, Accessible Volume, and Useful Accessibility}
\label{subsec:quantity_quality}

Global connectivity is a desirable property of a constrained mixer: without
it, portions of the feasible manifold may be permanently disconnected.
Connectivity is nevertheless an asymptotic structural statement. It does not
specify how much of that manifold is accessible after a specified finite
sequence of mixer rounds.

Even finite-depth accessible volume is insufficient to characterize the
optimization opportunity available to the circuit. Consider two schedules
$M$ and $M'$ satisfying
\begin{equation*}
    |R_p(M,x_0)|
    >
    |R_p(M',x_0)|.
\end{equation*}
No ordering of their best reachable objectives follows. It is entirely
possible that
\begin{equation}
    f_R^\star(M)
    <
    f_R^\star(M'),
    \label{eq:quantity_counterexample}
\end{equation}
because the smaller accessible region contains configurations of higher
objective quality.

Likewise, greater pair coverage does not imply a superior objective ceiling.
A schedule may activate many distinct feasible exchanges while steering the
accessible set toward configurations that remain poor under $f$. Another
schedule may repeatedly use a more selective set of transitions yet expose
the globally optimal solution within the same depth.

These observations separate three properties that are sometimes conflated:
\begin{equation*}
\boxed{
\begin{aligned}
    \text{connectivity}
    &\;\not\equiv\;
    \text{finite-depth accessible volume}\\
    &\;\not\equiv\;
    \text{reachable objective quality}.
\end{aligned}
}
\end{equation*}

This separation does not make $Q_R$ a complete model of QAOA performance.
It identifies structural opportunity: variational dynamics can redistribute
amplitude only among configurations made available by the executed schedule.
Interference, parameter optimization, and sampling determine how much of that
opportunity is realized.

The next section uses matched numbers of executed XY interactions to vary how
a limited transition budget is allocated and to examine its effects on pair
coverage, finite-depth accessible volume, reachable objective quality, and
realized QAOA performance.

\section{Resource-Controlled Mixer Design}
\label{sec:mixer_design}

Section~\ref{sec:accessibility} showed that finite-depth accessibility depends on both the available mixer family and the interactions actually scheduled in each layer. This raises a design question: given a limited number of constraint-preserving interactions, how should they be allocated across a shallow QAOA circuit?

We study this question using block-local XY mixers. XY interactions are a standard mechanism for preserving fixed-Hamming-weight constraints in the quantum alternating operator ansatz, and their connectivity and implementation structure have been studied extensively~\cite{hadfield2019alternating,wang2020xy,fuchs2022constrained}. Different XY topologies can produce substantially different circuit costs and algorithmic behavior, while the choice of initial state relative to the mixer can also affect constrained-QAOA performance~\cite{wang2020xy,He2023}. More recent work has further considered selective or adaptive allocation of mixer operations as a means of reducing circuit resources~\cite{ni2026adaptive}. Here, several controlled allocation policies are used to vary the finite-depth accessible set while holding the number of executed XY interactions fixed; selective mixing itself is not presented as a new primitive.

\subsection{Block-Local XY Mixing}
\label{subsec:block_xy}

Consider a feasible set defined by $G$ disjoint blocks,
\begin{equation*}
\mathcal{B}
=
\{\mathcal{B}_1,\ldots,\mathcal{B}_G\}
\end{equation*}
with block-cardinality constraints
\begin{equation}
\sum_{i\in \mathcal{B}_g} x_i = k_g,
\qquad
g=1,\ldots,G.
\label{eq:block_constraints}
\end{equation}
The corresponding feasible set is
\begin{equation}
\mathcal{F}
=
\left\{
x\in\{0,1\}^{n}:
\sum_{i\in \mathcal{B}_g}x_i=k_g
,\ \forall g
\right\}.
\label{eq:block_feasible_set}
\end{equation}

For each block, we permit exchange interactions only between variables belonging to that same block. An allowed pair $(i,j)$ therefore satisfies
\begin{equation*}
i,j\in \mathcal{B}_g
\end{equation*}
for some $g$. Its mixer generator is
\begin{equation}
H_{ij}^{XY}
=
X_iX_j+Y_iY_j.
\label{eq:block_xy_generator}
\end{equation}
Because this interaction exchanges occupations without changing their sum, it preserves the Hamming weight of $\mathcal{B}_g$. Products of such block-local exchanges consequently preserve every constraint in Eq.~\eqref{eq:block_constraints}. This construction follows the general use of XY-type mixers for Hamming-weight-constrained QAOA~\cite{hadfield2019alternating,wang2020xy}, while the restriction to disjoint interactions within an executed round provides a convenient resource-controlled setting for the finite-depth analysis developed here.

At layer $\ell$, the executed mixer round is
\begin{equation}
M_\ell
=
\bigcup_{g=1}^{G} M_{\ell,g},
\label{eq:block_round}
\end{equation}
where every $M_{\ell,g}$ contains only pairs from $\mathcal{B}_g$. We additionally require the pairs within an executed round to be vertex-disjoint. Hence no qubit participates in more than one XY exchange in the same round.

This restriction serves two purposes. First, the interactions in a round can be treated as simultaneously executable at the logical level. Second, it gives the mixer-layer transition semantics of Eq.~\eqref{eq:round_transition}: each active disjoint exchange can independently contribute an exchanged or unexchanged branch to the accessible basis-state set.

\subsection{Matched Mixer-Interaction Budget}
\label{subsec:mixer_budget}

Comparisons between mixer topologies can be misleading when one mixer simply executes more interactions than another. Complete, ring, parity, and other XY connectivity patterns have different implementation costs and different opportunities for state transfer~\cite{wang2020xy,fuchs2022constrained,He2023}. To separate interaction \emph{allocation} from interaction \emph{quantity}, we compare schedules using the same logical mixer budget whenever the policy has sufficient capacity.

For a depth-$p$ schedule
\begin{equation*}
M=(M_1,\ldots,M_p),
\end{equation*}
we define the executed XY-pair budget as
\begin{equation}
B
=
\sum_{\ell=1}^{p}
|M_\ell|.
\label{eq:mixer_budget}
\end{equation}
Thus $B$ counts the pair interactions that actually appear in the executed mixer
schedule, including repeated use of a pair in different layers. Throughout the
paper, the plain symbol $B$ denotes this interaction budget; calligraphic
$\mathcal{B}_g$ denotes constraint block $g$. The budget is distinct from the
number of \emph{unique} pairs covered by the schedule.

For a requested budget $B$, interactions are distributed as evenly as possible across the $p$ mixer rounds. In the depth-$p=3$ experiments considered below, the principal requested budgets correspond to the round quotas
\begin{equation}
\begin{aligned}
B=4  &: \quad (2,1,1), \\
B=8  &: \quad (3,3,2), \\
B=12 &: \quad (4,4,4), \\
B=16 &: \quad (6,5,5).
\end{aligned}
\label{eq:budget_quotas}
\end{equation}
Every schedule is validated directly from the executed pair list: all interactions must remain inside their constraint blocks, each round must be vertex-disjoint, and the total number of executed pairs must equal the intended budget unless the policy is explicitly identified as capacity-limited.

This accounting is defined at the logical-circuit level. Two schedules with the same $B$ can compile to different gate counts on a particular coupling graph. Section~\ref{sec:hardware} examines that distinction separately. Here, fixing $B$ allows us to ask a narrower question: when the same number of feasible exchange opportunities is available, does \emph{where and when} those interactions are placed change the quality of the finite-depth accessible region?

\subsection{Mixer Allocation Policies}
\label{subsec:mixer_policies}

We compare four allocation policies to create controlled variations of the same
transition budget. The policies range from static and coverage-oriented
schedules to an objective-informed schedule; they are not presented as four
separate claims of mixer novelty.

Let
\begin{equation*}
    E_{\rm allow}
    =
    \bigcup_{g=1}^{G}
    \left\{
        (i,j):
        i,j\in \mathcal{B}_g,\;
        i<j
    \right\}
\end{equation*}
denote the complete set of block-local exchange pairs. For policies whose
construction depends on the initial basis configuration, we further define
the initially active pair set
\begin{equation}
    A(x_0)
    =
    \left\{
        (i,j)\in E_{\rm allow}:
        x_{0,i}\neq x_{0,j}
    \right\}.
    \label{eq:initial_active_pairs}
\end{equation}
For a requested interaction budget $B$, let
$b_\ell$ denote the quota assigned to mixer round $\ell$, with
\begin{equation*}
    \sum_{\ell=1}^{p} b_\ell = B.
\end{equation*}
Each policy constructs a vertex-disjoint set $M_\ell$ satisfying
\begin{equation*}
    |M_\ell|\leq b_\ell.
\end{equation*}
Equality holds whenever the corresponding policy has sufficient pair
capacity.

\subsubsection{Fixed}

The \emph{fixed} policy begins from the initially active block-local pairs
$A(x_0)$ in a deterministic ordering: blocks are traversed in their fixed
order, and pairs within each block follow the ordering induced by
two-element combinations of the block indices. At every mixer layer, the
schedule greedily scans this same ordered list and accepts the first
vertex-disjoint pairs until the quota $b_\ell$ is filled or no further
compatible pair remains.

Equivalently, if $\mathcal{G}_{b_\ell}(C)$ denotes the greedy
vertex-disjoint prefix of an ordered pair list $C$ with capacity
$b_\ell$, then
\begin{equation}
    M_\ell^{\rm fixed}
    =
    \mathcal{G}_{b_\ell}\!\left(A(x_0)\right).
    \label{eq:fixed_schedule}
\end{equation}
The same ordered candidate set is reused at each layer, so individual pairs
may recur across mixer rounds. The policy provides a static
baseline in which executed interactions can be repeatedly concentrated on a
limited part of the available pair structure.

\subsubsection{Rotating-disjoint}

The \emph{rotating-disjoint} policy is constructed differently. Within each
block, the complete graph on the block variables is decomposed into a
deterministic sequence of round-robin disjoint matchings. The corresponding
blockwise matchings are combined to form a sequence
\begin{equation*}
    \mathcal{M}^{(1)},\ldots,\mathcal{M}^{(L)}
\end{equation*}
of globally vertex-disjoint candidate rounds. Mixer layer $\ell$ selects the
corresponding round cyclically and truncates it to the prescribed quota:
\begin{equation}
    M_\ell^{\rm rot}
    =
    \operatorname{prefix}_{b_\ell}
    \left(
        \mathcal{M}^{\,1+((\ell-1)\bmod L)}
    \right).
    \label{eq:rotating_schedule}
\end{equation}

Unlike the fixed, random-disjoint, and objective-improving constructions,
the rotating schedule is built from all block-local pairs, including pairs that
are inactive in $x_0$. Pairs connecting equal occupations can therefore
appear in the executed schedule; for a given basis configuration, such
interactions do not create an exchanged basis-state branch. The purpose of
the rotating construction is to distribute the available interaction budget
across different block-local pairings while maintaining vertex-disjointness
within every executed round.

\subsubsection{Random-disjoint}

The \emph{random-disjoint} policy also begins from the initially active pair
set $A(x_0)$. For each layer, a fresh copy of this candidate list is shuffled
using a seeded pseudorandom generator. The schedule then scans the shuffled
list and greedily accepts vertex-disjoint pairs until the quota $b_\ell$ is
filled or no additional compatible pair remains.

If $\pi_{\ell,s}$ denotes the permutation produced at layer $\ell$ by
schedule seed $s$, then
\begin{equation}
    M_\ell^{\rm rand}(s)
    =
    \mathcal{G}_{b_\ell}
    \left(
        \pi_{\ell,s}\!\left(A(x_0)\right)
    \right).
    \label{eq:random_schedule}
\end{equation}

This construction should not be interpreted as uniform sampling over all
possible disjoint matchings. Randomness enters through the shuffled edge
ordering, after which the matching is built greedily. The full active
candidate list is reshuffled independently at each layer, so pairs may be
reused across rounds. Different seeds generate different static schedules at
the same requested interaction budget.

\subsubsection{Objective-improving}

The first three policies do not rank candidate interactions according to
their local objective improvement. To construct a schedule that explicitly
targets objective-relevant accessibility, we additionally consider an
\emph{objective-improving} (OI) policy.

For a feasible initial configuration $x_0$, consider a legal within-block
exchange between an occupied variable $i$ and an unoccupied variable $j$,
oriented so that
\begin{equation*}
    x_{0,i}=1,
    \qquad
    x_{0,j}=0.
\end{equation*}
Let $x_0^{i\rightarrow j}$ denote the resulting feasible configuration obtained
by setting $x_i$ from $1$ to $0$ and $x_j$ from $0$ to $1$. We define the exact
objective change
\begin{equation}
    \Delta f_{i\rightarrow j}(x_0)
    =
    f\!\left(x_0^{i\rightarrow j}\right)-f(x_0).
    \label{eq:swap_delta_general}
\end{equation}
All legal held--unheld exchanges are scored once from the initial state
$x_0$. Candidates with
\begin{equation*}
    \Delta f_{i\rightarrow j}(x_0)>0
\end{equation*}
are retained and ordered by decreasing objective gain, with the pair indices
used as a deterministic tie-break.

Let
\begin{equation*}
    C_{\rm OI}(x_0)
    =
    \operatorname{sort}_{\Delta f\downarrow}
    \left\{
        (i,j)\in A(x_0):
        \Delta f_{i\rightarrow j}(x_0)>0
    \right\}.
\end{equation*}
The executed OI round is then
\begin{equation}
    M_\ell^{\rm OI}
    =
    \mathcal{G}_{b_\ell}
    \left(
        C_{\rm OI}(x_0)
    \right).
    \label{eq:oi_schedule}
\end{equation}

The ranking is fixed when the schedule is constructed. Objective gains are
not recomputed after an accepted pair and are not updated between QAOA
layers. Likewise, previously selected pairs are not penalized or excluded in
later layers, so the same high-ranked interaction may be reused across
rounds. The OI policy is therefore objective-informed but not adaptive during
circuit execution.

For the transaction-aware quadratic portfolio objective used in
Sec.~\ref{sec:experiments},
\begin{equation}
    f(x)
    =
    \mu^\top x
    -
    q x^\top\Sigma x
    -
    c_{\rm tc}\lVert x-r\rVert_1,
    \label{eq:portfolio_objective_preview}
\end{equation}
the exact change associated with exchanging a held asset $i$ for an unheld
asset $j$, evaluated from $x_0$, is
\begin{align}
    \Delta f_{i\rightarrow j}
    ={}&
    (\mu_j-\mu_i)
    \nonumber\\
    &-
    q\left[
        (\Sigma_{jj}-\Sigma_{ii})
        +
        2\sum_{k\neq i,j}
        (\Sigma_{jk}-\Sigma_{ik})x_{0,k}
    \right]
    \nonumber\\
    &-
    c_{\rm tc}
    \left(
        \left\lVert x_0^{i\rightarrow j}-r\right\rVert_1
        -
        \left\lVert x_0-r\right\rVert_1
    \right).
    \label{eq:exact_swap_delta}
\end{align}
The transaction-cost reference $r$ is fixed by the problem definition and is
not changed when the QAOA initial state is varied. Consequently, in the
warm-start experiments the state used to score candidate exchanges, $x_0$,
and the fixed transaction-cost reference $r$ can differ.

Objective-aware or adaptive selection of mixer operations is not unique to
this work; recent adaptive mixer-allocation approaches also use
problem-dependent evaluation criteria to decide which mixer operations to
apply~\cite{ni2026adaptive}. The OI construction used here has a narrower
role. It is a static, objective-informed schedule constructor used to test
whether directing a fixed interaction budget toward locally improving
transitions changes the objective quality contained in the finite-depth
accessible set.

We do not assume that OI is optimal or that positive single-swap gains identify
the best mixer schedule in general. A static local ranking can miss
transitions that become useful only after other exchanges occur. In this
study, OI is a controlled intervention on the objective content of the
accessible region; it is not an adaptive optimization procedure.

\subsection{Capacity-Limited Schedules}
\label{subsec:capacity_limited}

A distinction is required between a \emph{requested} mixer budget and the number of interactions a policy can actually execute. Fixed, rotating, and random schedules can generally fill the prescribed quotas whenever sufficient valid disjoint pairs exist. OI is more restrictive because it accepts only positive-gain exchanges.

For some initial states or problem instances, fewer than $B$ mutually compatible positive-gain interactions may be available across the scheduled rounds. We define the resulting effective budget as
\begin{equation}
B_{\rm eff}
=
\sum_{\ell=1}^{p}|M_\ell|
\leq B .
\label{eq:effective_budget}
\end{equation}
If $B_{\rm eff}<B$, the run is labeled \emph{capacity-limited} and is not silently treated as resource matched.

This accounting prevents resource mismatch from being mistaken for a schedule-quality effect. Principal matched-budget comparisons use schedules satisfying
\begin{equation*}
B_{\rm eff}=B,
\end{equation*}
while capacity-limited cases are reported separately as diagnostics of the objective-informed policy.

\subsection{Structural Diagnostics}
\label{subsec:structural_diagnostics}

Each constructed schedule is characterized before QAOA parameter optimization. Besides the finite-depth quantities introduced in Sec.~\ref{sec:accessibility}, we record the number of unique allowed pairs appearing anywhere in the schedule. If
\begin{equation*}
E_{\rm sched}
=
\bigcup_{\ell=1}^{p}M_\ell
\end{equation*}
and $E_{\rm allow}$ denotes all block-local pairs permitted by the constraints, the pair-coverage fraction is
\begin{equation}
C_{\rm pair}
=
\frac{|E_{\rm sched}|}{|E_{\rm allow}|}.
\label{eq:pair_coverage}
\end{equation}

We then evaluate four complementary properties:
\begin{equation*}
\begin{aligned}
C_{\rm pair}
&=
\frac{|E_{\rm sched}|}{|E_{\rm allow}|},
&
A_R
&=
\frac{|R_p|}{|\mathcal{F}|},
\\
D_R
&=
\frac{\left|\{x\in R_p:f(x)>f(x_0)\}\right|}{|R_p|},
&
Q_R
&=
\frac{f_R^\star-f(x_0)}{f^\star-f(x_0)}.
\end{aligned}
\end{equation*}

These diagnostics deliberately separate progressively stronger notions of mixer opportunity. Pair coverage asks how broadly the schedule uses the allowed interaction graph. $A_R$ asks how much of the feasible configuration set those interactions make structurally accessible at the executed depth. $D_R$ asks whether the accessible region is enriched in configurations that improve on the initial state. Finally, $Q_R$ asks whether that region contains a high-quality solution at all.

The central empirical question is how strongly each diagnostic is associated with realized shallow-QAOA performance, not whether one policy maximizes every diagnostic. In particular, the framework of Sec.~\ref{sec:accessibility} predicts a qualitative distinction between exposing \emph{more} feasible configurations and exposing \emph{better} feasible configurations. The resource-controlled policies above allow that distinction to be tested without conflating it with the total number of executed mixer interactions.

\section{Experimental Design and Portfolio Benchmark}
\label{sec:experiments}

The accessibility framework of Sec.~\ref{sec:accessibility} is not specific to
portfolio optimization, but portfolio selection provides a useful test bed:
the objective contains dense return--risk interactions, practical allocation
rules naturally induce cardinality constraints, and portfolio updates can be
measured relative to a previous allocation. Classical portfolio optimization
provides the standard language for expected return, risk, allocation
constraints, and implementation costs~\cite{Markowitz1952,CornuejolsTutuncu2007}.

Quantum approaches to portfolio optimization have been studied using the
quantum alternating operator ansatz, standard and constrained QAOA,
warm-starting, and other variational constructions
\cite{Hodson2019,Mugel2022,Brandhofer2023,He2023,kordonowy2026lie}.
Recent work continues to investigate cardinality-preserving XY mixers,
initialization strategies, and alternative constrained ans\"atze for portfolio
problems \cite{Mancilla2026,Yamamura2026,Falla2026}. We use a market-grounded,
controlled portfolio benchmark to test whether finite-depth accessible
\emph{quality} provides information about shallow-QAOA performance beyond the
size of the accessible set itself.

\subsection{Historical Equity Benchmark}
\label{subsec:portfolio_data}

The benchmark is constructed from a historical U.S.-equity price matrix
covering the period from 2013-01-02 to 2023-01-30, with 2537 daily price
observations. The source matrix contains 121 ticker columns. The executable
data pipeline explicitly drops \texttt{GOOGL}; the available provenance does
not record why this ticker was excluded. After this fixed preprocessing step,
120 equity price series remain before portfolio-universe screening.

The original data matrix predates the present research repository, and its
original market-data provider and price-field convention are not recoverable
from the available provenance records. Accordingly, we make no provider-specific
or index-membership claim and treat the matrix simply as the fixed historical
equity input to the benchmark.

A fixed company-sector mapping is used to impose economically interpretable
block structure. Within this fixed input universe and fixed sector map, asset
screening uses only the first 63 trading days, spanning 2013-01-02 to
2013-04-03, before the return and covariance parameters used in optimization
are estimated. The five sectors with the largest numbers of mapped assets are
retained. Within each of these sectors, four candidate assets are chosen by
descending mean price over the 63-day screening window. This ranking is used
only as a deterministic universe-construction rule; it is not interpreted as
a return forecast or investment signal.

The resulting optimization universe is shown in
Table~\ref{tab:portfolio_universe}.

\begin{table*}[t]
\caption{
Selected portfolio universe. The benchmark contains five sector blocks with
four candidate assets per block and an exact quota of two selected assets per
block.
}
\label{tab:portfolio_universe}
\begin{ruledtabular}
\begin{tabular}{l l c}
Sector & Candidate assets & Quota \\
\hline
Health Care
    & REGN, BDX, TMO, HUM
    & 2 \\
Information Technology
    & IBM, ACN, INTU, QCOM
    & 2 \\
Financials
    & BLK, GS, AON, TROW
    & 2 \\
Industrials
    & GE, CMI, DE, MMM
    & 2 \\
Consumer Discretionary
    & BKNG, ORLY, HD, TGT
    & 2 \\
\end{tabular}
\end{ruledtabular}
\end{table*}

Thus the canonical portfolio contains $n=20$ binary decision variables with
total cardinality $\sum_{i=1}^{20}x_i=10$. More specifically, for the five
sector blocks $\mathcal{B}_1,\ldots,\mathcal{B}_5$,
\begin{equation}
    \sum_{i\in \mathcal{B}_g}x_i=2,
    \qquad
    g=1,\ldots,5.
    \label{eq:portfolio_sector_constraints}
\end{equation}
The feasible-set size is therefore
\begin{equation}
    |\mathcal{F}|
    =
    \prod_{g=1}^{5}
    \binom{4}{2}
    =
    6^5
    =
    7776.
    \label{eq:portfolio_feasible_size}
\end{equation}

The exact block constraints have a dual role. Financially, they impose a
stylized form of sector diversification. Algorithmically, they create a
factorized feasible manifold in which every allowed XY exchange has a clear
interpretation as replacing one selected asset by one unselected asset within
the same sector. The benchmark is market-grounded but intentionally
stylized: it studies discrete asset selection rather than continuous portfolio
weights, and does not attempt to model the full set of liquidity, factor,
regulatory, tax, or market-impact constraints encountered in institutional
portfolio management.

\subsection{Return, Risk, and Transaction-Cost Model}
\label{subsec:portfolio_objective}

Daily simple returns are computed from the selected price series. The 2536
resulting return observations are divided into a training period
(2013-01-03 to 2021-01-22; 2028 observations) and a held-out period
(2021-01-25 to 2023-01-30; 508 observations). All coefficients entering the optimization objective are estimated from the
training period. The held-out period is retained as a data-separation
boundary; the QAOA results reported here should not be interpreted as an
out-of-sample investment-performance backtest.

The portfolio objective follows the classical return--risk trade-off introduced
by mean--variance portfolio theory \cite{Markowitz1952}, with additional
turnover cost. Because sample expected returns can be unstable, the annualized
training-period mean
\begin{equation*}
    \widehat{\mu}_{\rm samp}
    =
    252\,\overline{r}
\end{equation*}
is shrunk toward its cross-sectional mean,
\begin{equation}
    \widehat{\mu}_{\rm sh}
    =
    (1-\lambda_\mu)
    \widehat{\mu}_{\rm samp}
    +
    \lambda_\mu
    \overline{\mu}_{\rm samp}\mathbf{1},
    \qquad
    \lambda_\mu=0.5.
    \label{eq:mean_shrinkage}
\end{equation}
This use of shrinkage is motivated by the established role of
Stein/Bayes--Stein-type estimators in stabilizing expected-return estimates
for portfolio construction \cite{Jorion1986}. We do not require
Eq.~\eqref{eq:mean_shrinkage} to coincide with a particular empirical
Bayes estimator; it is the fixed shrinkage rule used throughout the benchmark.

Risk is estimated using the Ledoit--Wolf shrinkage covariance estimator
\cite{LedoitWolf2004}. If
$\widehat{\Sigma}_{\rm LW}^{\rm daily}$ denotes the covariance returned from
the training-period daily returns, the annualized estimate is
\begin{equation}
    \widehat{\Sigma}_{\rm LW}
    =
    252\,
    \widehat{\Sigma}_{\rm LW}^{\rm daily}.
    \label{eq:lw_annualized}
\end{equation}

Before constructing the optimization Hamiltonian, the return and covariance
terms are independently scaled:
\begin{align}
    s_\mu
    &=
    \max_i
    \left|
        \widehat{\mu}_{{\rm sh},i}
    \right|,
    &
    \mu
    &=
    \frac{\widehat{\mu}_{\rm sh}}{s_\mu},
    \label{eq:mu_normalization}
    \\
    s_\Sigma
    &=
    \max_{i,j}
    \left|
        \widehat{\Sigma}_{{\rm LW},ij}
    \right|,
    &
    \Sigma
    &=
    \frac{\widehat{\Sigma}_{\rm LW}}{s_\Sigma}.
    \label{eq:sigma_normalization}
\end{align}

For a feasible binary portfolio $x$, the objective maximized throughout the
portfolio experiments is
\begin{equation}
    f(x)
    =
    \mu^{T}x
    -
    q\,x^{T}\Sigma x
    -
    c_{\rm tc}
    \lVert x-r\rVert_1,
    \qquad
    q=1.
    \label{eq:portfolio_objective}
\end{equation}
The first two terms represent normalized expected return and covariance risk,
respectively. The final term penalizes turnover relative to a fixed reference
portfolio $r$. Transaction-aware portfolio formulations have also appeared in
quantum portfolio-rebalancing studies
\cite{Hodson2019,Mugel2022}; here the term is deliberately simple so that its
effect on the discrete objective remains transparent.

The nominal trading-cost scale is 10 basis points per rebalance with an
annualization factor of four,
\begin{equation*}
    c_{\rm tc}^{\rm ann}
    =
    4
    \left(
        \frac{10}{10^{4}}
    \right)
    =
    0.004.
\end{equation*}
It is placed on the same return normalization scale through
\begin{equation}
    c_{\rm tc}
    =
    \frac{c_{\rm tc}^{\rm ann}}{s_\mu}.
    \label{eq:tc_normalization}
\end{equation}
For the canonical $n=20$ instance this gives
\begin{equation*}
    c_{\rm tc}
    =
    0.0168522.
\end{equation*}

The reference portfolio $r$ is the deterministic canonical incumbent obtained
by selecting the first two candidates in each sector block. It serves as a
fixed objective reference; it is not intended to represent a historical investor portfolio.
That separation matters in the warm-start study of
Sec.~\ref{sec:robustness}: changing the QAOA initialization $x_0$ does
\emph{not} change $r$. Thus initial-state effects are not confounded with a
change in the transaction-cost objective.

The quadratic risk term generally induces a dense cost Hamiltonian. Under the
standard binary-to-Ising transformation, Eq.~\eqref{eq:portfolio_objective}
gives a diagonal phase-separation Hamiltonian of the form
\begin{equation}
    H_C
    =
    \sum_i h_i Z_i
    +
    \sum_{i<j}J_{ij}Z_iZ_j
    +
    C,
    \label{eq:portfolio_ising}
\end{equation}
where $C$ is an irrelevant constant. The transaction term contributes only
to the one-body fields because $r$ is fixed. This dense phase separator is
also important for the compilation analysis in Sec.~\ref{sec:hardware},
where cost-layer and mixer-layer resources are considered separately.

\subsection{QAOA Simulation and Parameter Optimization}
\label{subsec:qaoa_protocol}

The principal experiments use shallow QAOA at depth $p=3$, consistent with the finite-depth regime that motivates the present analysis.
QAOA was originally formulated as a depth-indexed variational algorithm for
combinatorial optimization \cite{Farhi2014}; constrained variants at similarly
small depths have subsequently been studied in portfolio settings
\cite{He2023,Brandhofer2023}.

All primary QAOA experiments are performed using noiseless shot-based
simulation with Qiskit Aer. This choice is deliberate. The purpose of the
main experiment is to isolate differences induced by the executed mixer
schedule without combining them with calibration drift, device noise, or
hardware-routing effects. Physical connectivity and compilation are examined
separately in Sec.~\ref{sec:hardware}.

For every mixer policy, the variational parameters are trained using the
Conditional Value-at-Risk (CVaR) objective with $\alpha=0.10$.
CVaR optimization has been proposed for variational combinatorial optimization
as a way to emphasize favorable low-energy samples relative to optimizing the
expectation value alone \cite{Barkoutsos2020}. Because the financial objective $f(x)$ is maximized, the QAOA implementation
uses the energy convention $E(x)=-f(x)$.
The training loss therefore averages the lowest-energy $\alpha$ fraction of
the sampled distribution, corresponding to the highest-objective tail under
this sign convention.

Each optimizer evaluation uses 2048 circuit samples. Parameters are optimized
with COBYLA for at most 300 iterations and four independent restarts. The
first restart is initialized at
\begin{equation*}
    \gamma_\ell=0.02,
    \qquad
    \beta_\ell=0.20,
    \qquad
    \ell=1,\ldots,p,
\end{equation*}
while subsequent restarts perturb this initialization with Gaussian jitter of
scale $0.20$. Warm-starting variational optimization is well established as a way to improve
low-depth optimization behavior~\cite{Egger2021,He2023}. A common
parameter-initialization rule is used across mixer policies so that the
initialization does not create a schedule-specific advantage.

In addition, a global angle-scale parameter
$s\in\{0.8,1.0,1.4\}$ is evaluated. At circuit construction, both phase-separator and mixer angles
are multiplied by $s$. For each configuration, the run with the lowest
training CVaR is retained. Once the optimized parameters are selected, the
corresponding circuit is sampled using
\begin{equation*}
    N_{\rm final}=8192
\end{equation*}
shots for final evaluation.

All compared constrained schedules use the same depth, training rule, shot
counts, objective, and parameter-optimization protocol. The schedule itself
and, where explicitly varied, the feasible initial basis state are therefore
the principal controlled algorithmic differences.

\subsection{Exact Structural References and Evaluation Metrics}
\label{subsec:evaluation_metrics}

The canonical feasible set contains only 7776 configurations, which permits
exact enumeration of the structural reference quantities used in this work.
For each portfolio instance, we evaluate the complete feasible objective
landscape and compute
\begin{equation}
    f^\star
    =
    \max_{x\in\mathcal{F}}f(x).
    \label{eq:experimental_global_optimum}
\end{equation}
For every executed mixer schedule, the recursion in
Sec.~\ref{subsec:round_accessibility} is applied independently of the
variational optimization to obtain $R_p(M,x_0)$, followed by the structural
quantities $|R_p|$, $f_R^\star$, and $Q_R$, together with the pair-coverage and improving-state diagnostics defined in
Sec.~\ref{subsec:structural_diagnostics}.

After QAOA training, $f_{\rm best}$ denotes the best feasible objective value
observed among the final samples. The implementation reports the normalized
realized score
\begin{equation}
    S_{\rm eff}
    =
    \frac{f_{\rm best}-f_{\rm ref}}{f^\star-f_{\rm ref}},
    \label{eq:effective_score}
\end{equation}
where $f_{\rm ref}$ is the fixed reporting reference used for the corresponding
experiment. In the canonical portfolio experiment,
$f_{\rm ref}=f(x_0)$ because the initial state is the canonical incumbent. In
the warm-start study, $x_0$ varies while the reporting reference remains fixed
at the canonical incumbent. Thus $S_{\rm eff}=0$ means that the best sampled
objective equals the reporting reference and $S_{\rm eff}=1$ means that a
globally optimal feasible configuration was sampled.

Accordingly, $Q_R$ measures structural opportunity from the actual initial
state, whereas $S_{\rm eff}$ reports realized best-sample performance using
the experiment's reporting reference.
The support result of Sec.~\ref{subsec:support_containment} imposes a structural
ceiling through $f_R^\star$, but it does not guarantee that variational
optimization concentrates amplitude near that ceiling. CVaR optimization,
interference, the parameter landscape, and finite-shot sampling all operate
\emph{within} the opportunity permitted by the executed schedule.

Feasibility is checked directly on the sampled bit strings by verifying every
block constraint in Eq.~\eqref{eq:portfolio_sector_constraints}. For the
ideal block-local XY circuits this serves primarily as an implementation
consistency check, since feasibility preservation is already guaranteed by
the mixer construction.

\subsection{Experimental Hierarchy}
\label{subsec:experimental_hierarchy}

The empirical study is organized to separate mechanism from robustness. We
first use the canonical $n=20$ portfolio instance to compare the four mixer
policies of Sec.~\ref{subsec:mixer_policies} under matched executed-interaction
budgets $B\in\{8,16\}$. This experiment asks whether schedules with comparable logical mixer resources
can expose accessible regions of substantially different objective quality,
and whether those differences are reflected in realized shallow-QAOA
performance.

We then vary the feasible initial state while keeping the transaction-cost
reference $r$ fixed, and construct nested portfolio instances with
$n=14$, $16$, and $20$. These tests examine whether the observed accessibility
mechanism depends on a particular warm start or problem size. Finally, the
same accessibility diagnostics are applied to a synthetic block-constrained
QUBO family, which removes the financial structure of the objective while
retaining the block-cardinality constraint structure. These robustness and
generalization tests are presented in Sec.~\ref{sec:robustness}.

The next section asks a focused empirical question: under controlled
shallow-circuit resources, does the \emph{quality} of the finite-depth
accessible region provide a more informative description of realized QAOA
behavior than the number of pairs or feasible configurations exposed by the
schedule?

\section{Reachable Quantity Versus Reachable Quality}
\label{sec:results}

Section~\ref{sec:accessibility} separates the number of feasible configurations
made accessible at finite depth from the objective quality contained in that
set. These two properties can differ substantially; this section tests that
claim empirically.

The first experiment uses the canonical $n=20$ portfolio instance of
Sec.~\ref{sec:experiments}. All mixer policies are compared at the same QAOA
depth, under the same optimization protocol, and at matched executed XY
interaction budgets. The comparison asks whether different allocations of the
same finite transition budget expose regions of the feasible manifold with
different optimization value, and whether those structural differences are
reflected in realized shallow-QAOA performance. It is not intended as a
universal ranking of mixer policies.

\subsection{Canonical Resource-Matched Comparison}
\label{subsec:canonical_results}

We first compare the fixed, rotating-disjoint, random-disjoint, and
objective-improving schedules at $B\in\{8,16\}$. 
For random-disjoint schedules, five independently generated schedules are
evaluated at each budget; the other three policies are deterministic for the
canonical configuration.

Table~\ref{tab:canonical_results} reports the realized effective score
$S_{\rm eff}$ after optimization with favorable-tail CVaR. The random-disjoint
entries are averages over the five schedule seeds.

\begin{table}[tbh]
\caption{
Realized effective QAOA score on the canonical $n=20$ portfolio instance.
All schedules use $p=3$ and are compared at matched numbers of executed XY
pair interactions. Random-disjoint values are means over five independently
generated schedules. Larger $S_{\rm eff}$ is better.
}
\label{tab:canonical_results}
\begin{ruledtabular}
\begin{tabular}{lcc}
Mixer policy & $B=8$ & $B=16$ \\
\hline
Fixed                  & 0.0788 & 0.4577 \\
Rotating-disjoint      & 0.0788 & 0.4577 \\
Random-disjoint        & 0.6872 & 0.9246 \\
Objective-improving    & 0.8896 & 1.0000 \\
\end{tabular}
\end{ruledtabular}
\end{table}

The comparison reveals a substantial dependence on how the same logical mixer
budget is allocated. At $B=8$, the fixed and rotating schedules attain only a
small fraction of the available improvement, whereas alternative random
allocations produce substantially larger scores on average. The objective-improving schedule attains $S_{\rm eff}=0.8896$. At $B=16$,
random-disjoint schedules reach a mean effective score of $0.9246$, while the
objective-improving schedule samples a globally optimal feasible configuration
and reaches $S_{\rm eff}=1$.
\begin{figure*}[t]
    \centering
    \includegraphics[width=0.8\linewidth]{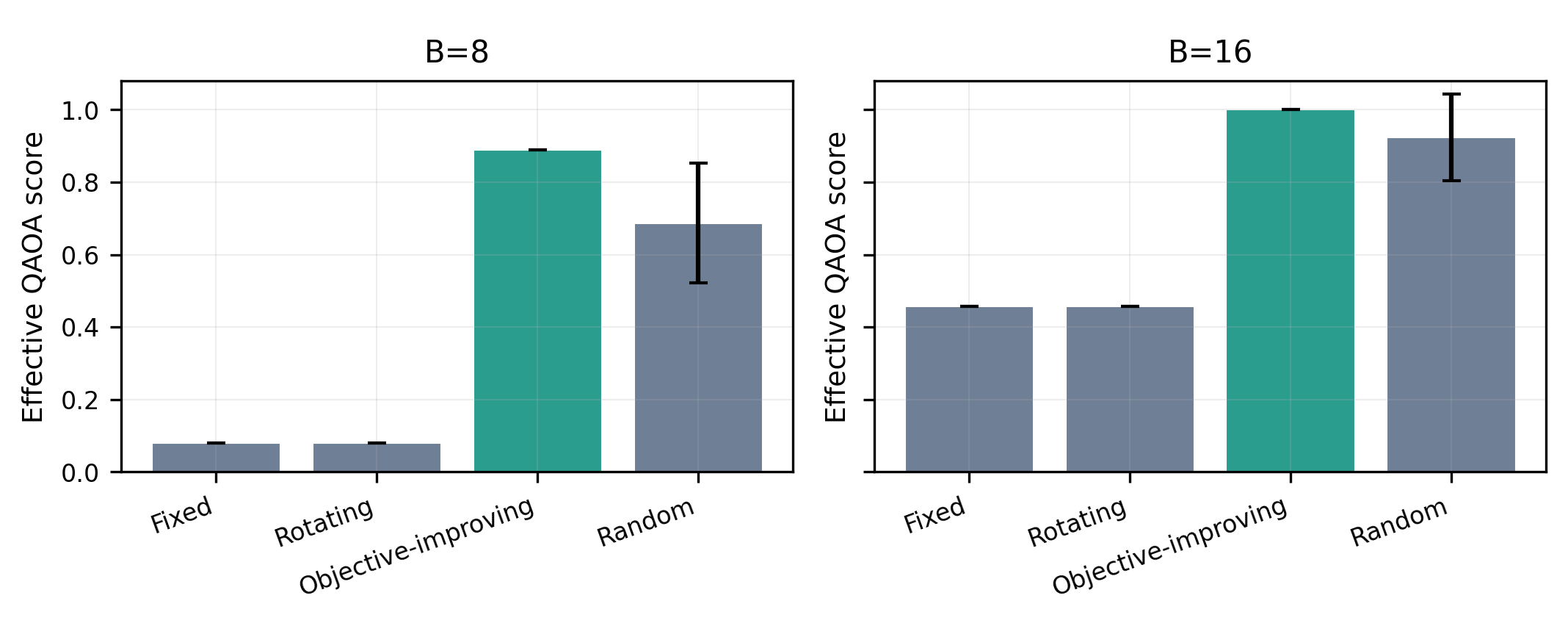}
    \caption{Canonical resource-matched QAOA performance at requested XY-interaction
    budgets $B=8$ and $B=16$. Bars show the effective QAOA score for the
    deterministic mixer policies and the mean score across random-disjoint
    schedule seeds; error bars indicate the corresponding variability where
    applicable. Despite matching the requested interaction budget, the mixer
    allocations produce substantially different shallow-QAOA performance.}
    \label{fig:canonical_effective_score}
\end{figure*}
These observations should not be interpreted as establishing universal
superiority of objective-informed scheduling. Random schedules are already
competitive at the larger budget, and Sec.~\ref{sec:robustness} identifies
settings in which objective-improving schedules become capacity limited.
Instead, the canonical experiment provides a controlled example in which
schedules using the same number of mixer interactions expose substantially
different optimization opportunities.

\subsection{Accessible Volume Is Not Accessible Quality}
\label{subsec:canonical_quantity_quality}

The performance differences in
Table~\ref{tab:canonical_results} cannot be interpreted solely through the
number of distinct mixer pairs or the size of the finite-depth accessible set.
For each executed schedule we independently compute
\begin{equation*}
    \frac{|R_p|}{|\mathcal{F}|},
    \qquad
    \frac{|\{x\in R_p:f(x)>f(x_0)\}|}{|R_p|},
    \qquad
    Q_R
\end{equation*}
before using the QAOA samples themselves.

These quantities answer different questions. Pair coverage measures the
diversity of local exchanges exposed by the schedule. Reachable fraction
measures how much of the full feasible manifold can be generated after the
executed mixer rounds. In contrast, $Q_R$ asks whether that accessible region
contains a solution that closes a substantial fraction of the objective gap
between $x_0$ and the global feasible optimum.

Figure~\ref{fig:canonical_effective_score} illustrates that these quantities do not collapse to a single notion of ``more exploration.'' Schedules with broad
structural exposure can access many feasible configurations without
necessarily exposing the best objective values available at that depth.
Conversely, a more selective schedule may expose fewer configurations while
including substantially higher-quality states.

The $B=16$ objective-improving schedule provides an especially clear example.
Because the globally optimal feasible configuration belongs to its
finite-depth accessible set, $Q_R=1$, and the optimized circuit subsequently
samples that optimum, giving $S_{\rm eff}=1$.
The equality of these two quantities in this particular run should not be
confused with a general identity. The structural result of
Sec.~\ref{subsec:support_containment} instead implies the unnormalized
best-sample ceiling
\begin{equation*}
    f_{\rm best}\leq f_R^\star.
\end{equation*}
If realized and reachable scores are normalized using the same reference
floor, the corresponding normalized inequality follows. In the warm-start
comparisons of Sec.~\ref{sec:robustness}, however, the reported
$S_{\rm eff}$ and $Q_R$ use different reference normalizations. We therefore
interpret $Q_R$ as a structural quality diagnostic rather than as a rowwise
upper bound on the reported $S_{\rm eff}$.


The canonical results support the distinction developed in
Sec.~\ref{subsec:quantity_quality}:
\begin{equation*}
\boxed{
\begin{aligned}
    \text{accessible quantity}
    &\;\not\equiv\;
    \text{accessible objective quality}\\
    &\;\not\equiv\;
    \text{realized QAOA performance}.
\end{aligned}
}
\end{equation*}

\subsection{Which Accessibility Metric Tracks Realized QAOA?}
\label{subsec:mechanistic_association}

The canonical experiment motivates a broader mechanistic question: across
different schedules and problem settings, which finite-depth accessibility
quantity shows the strongest association with the realized QAOA score?

To address this question, we harmonize the completed QAOA configurations from
the canonical portfolio, warm-start variants, nested portfolio instances, and
synthetic block-constrained QUBOs. The detailed robustness results are
presented in Sec.~\ref{sec:robustness}; here we use the combined data only to
compare the accessibility diagnostics against $S_{\rm eff}$.

The data contain $N=207$ completed QAOA configurations. Table~\ref{tab:accessibility_associations}
reports pooled Pearson and Spearman associations together with a partial
Pearson association controlling for problem size, executed XY budget, and
problem domain.

\begin{table*}[t]
\caption{
Association between finite-depth accessibility diagnostics and realized
effective QAOA score across the 207 completed configurations. The partial
association controls for problem size $n$, the number of executed XY
interactions, and domain fixed effects. These statistics are descriptive and
should not be interpreted as independent-sample causal estimates because
multiple schedule rows arise from the same underlying problem instances.
}
\label{tab:accessibility_associations}
\begin{ruledtabular}
\begin{tabular}{lccc}
Accessibility diagnostic
&
Pearson
&
Spearman
&
Partial Pearson
\\
\hline
Unique-pair coverage
    & 0.2657 & 0.2778 & 0.0465 \\
Reachable fraction $|R_p|/|\mathcal{F}|$
    & 0.3829 & 0.4170 & 0.3064 \\
Fraction of $R_p$ improving on $x_0$
    & 0.2104 & 0.2556 & 0.2375 \\
Best reachable normalized quality $Q_R$
    & \textbf{0.7003} & \textbf{0.8057} & \textbf{0.6996} \\
\end{tabular}
\end{ruledtabular}
\end{table*}

The strongest association is obtained for the best reachable normalized
quality. In the pooled data,
\begin{equation}
    \rho_{\rm P}(Q_R,S_{\rm eff})=0.7003,
    \qquad
    \rho_{\rm S}(Q_R,S_{\rm eff})=0.8057 ,
    \label{eq:QR_pooled_correlations}
\end{equation}
and the partial association remains
\begin{equation}
    \rho_{\rm partial}=0.6996
    \label{eq:QR_partial}
\end{equation}
after controlling for $n$, executed XY budget, and domain.

Reachable fraction also exhibits a positive relationship with realized
performance, with pooled Pearson correlation $0.3829$ and partial association
$0.3064$. The corrected results do not support the stronger statement that
accessible volume is uninformative. Reachable fraction carries some
performance-relevant information, although its association is substantially
weaker here than that of the best objective quality contained in the
accessible region.

\begin{figure*}[t]
    \centering
    \includegraphics[width=0.95\textwidth]
    {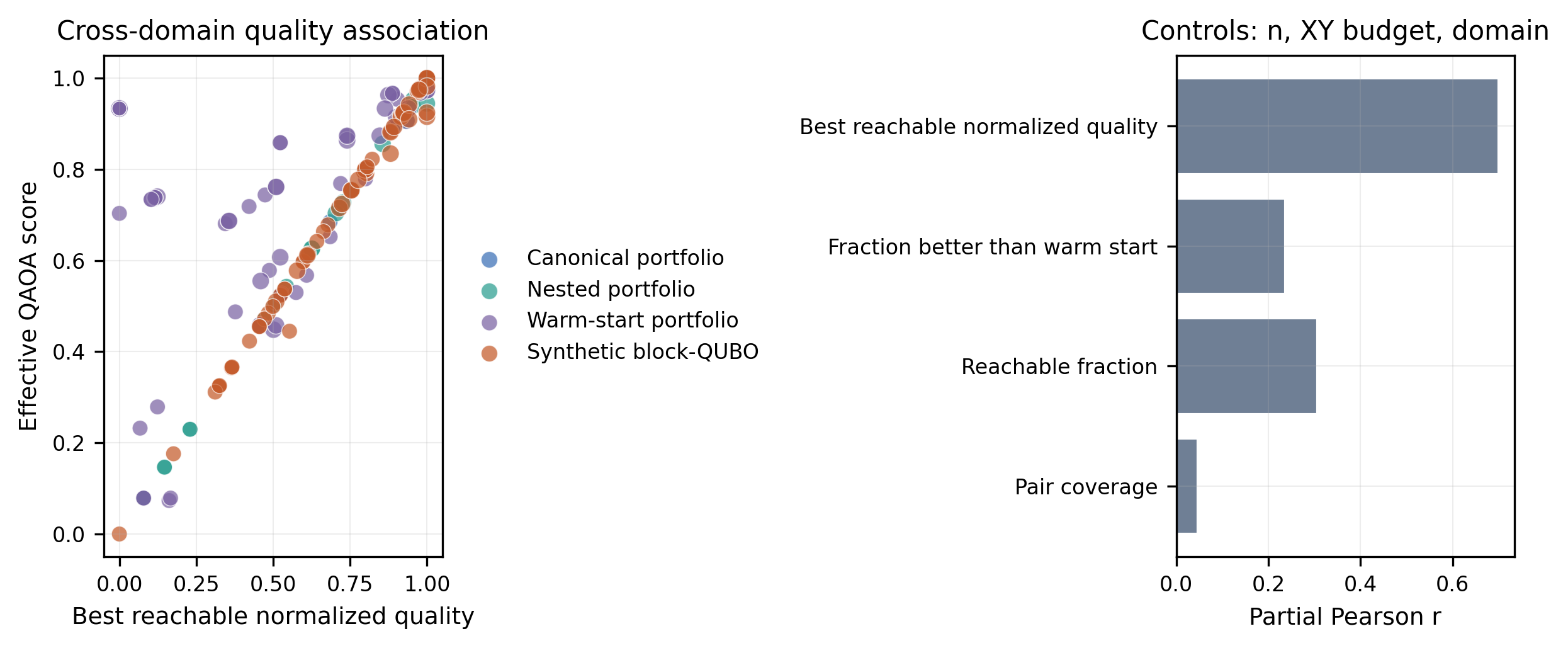}
    \caption{
    Cross-domain relationship between finite-depth reachable quality and
    realized shallow-QAOA performance. Left: effective QAOA score
    $S_{\rm eff}$ versus best reachable normalized quality $Q_R$ across the
    207 completed portfolio and synthetic QAOA configurations. Right: partial
    Pearson associations between $S_{\rm eff}$ and the structural diagnostics
    after controlling for problem size $n$, executed XY-interaction budget,
    and problem domain. Among the tested diagnostics, $Q_R$ shows the strongest
    controlled association with realized performance. The associations are
    observational rather than causal.
    }
    \label{fig:cross_domain_quality}
\end{figure*}

Pair coverage shows a still weaker mechanistic relationship. Although its
pooled Pearson correlation is positive, the association falls to $0.0465$
after controlling for size, executed mixer budget, and domain. This is
consistent with the structural argument of
Sec.~\ref{subsec:quantity_quality}: exposing a larger number of distinct
local exchanges does not specify which objective values those exchanges make
accessible within the available depth.
Figure~\ref{fig:cross_domain_quality} summarizes these relationships across
the harmonized portfolio and synthetic experiments.

\subsection{Dependence-Aware Robustness of the Association}
\label{subsec:cluster_association}

The 207 rows are not statistically independent: multiple mixer schedules,
budgets, and initial states can correspond to the same underlying optimization
instance. Treating every row as an independent observation would therefore
overstate the effective sample size. We account for this dependence by
clustering observations at the problem-instance level
\cite{CameronMiller2015}.

The resulting analysis contains 16 problem-instance clusters. For $Q_R$, the
cluster-robust controlled coefficient is
\begin{equation}
    \widehat{\beta}_{Q_R}
    =
    0.5893,
    \qquad
    {\rm SE}=0.1673,
    \qquad
    p=0.0031.
    \label{eq:QR_cluster}
\end{equation}
A cluster bootstrap gives
\begin{equation}
    \widehat{\beta}_{Q_R}^{\rm boot}
    =
    0.6202,
    \qquad
    95\%~{\rm CI}
    =
    [0.3479,\,0.9761].
    \label{eq:QR_bootstrap}
\end{equation}

Reachable fraction also remains positively associated with realized score in
this dependence-aware analysis,
\begin{equation*}
    \widehat{\beta}_{|R_p|}
    =
    0.4584,
    \qquad
    p=0.0003,
\end{equation*}
with cluster-bootstrap interval
\begin{equation*}
    [0.2863,\,0.7063].
\end{equation*}
In contrast, the cluster-robust coefficient for unique-pair coverage is
$0.0963$ with $p=0.1349$, and its bootstrap interval includes zero.

These dependence-aware results reinforce the structural interpretation.
Reachable volume matters in the tested data, but the strongest and most
consistent accessibility signal is the objective quality contained within the
finite-depth accessible region. In particular, $Q_R$ remains the largest
association under pooled, rank-based, controlled, cluster-robust, and
cluster-bootstrap analyses.

\subsection{Structural Opportunity, Not a Performance Guarantee}
\label{subsec:results_interpretation}

The statistical association should not be read as a causal or universal
prediction rule. The quantity $Q_R$ is computed entirely from the structural
accessible set and the classical objective landscape; it contains no
information about the optimized QAOA angles or the probability assigned to
individual configurations.

Accordingly, a large value of $Q_R$ is best interpreted as
\emph{structural opportunity}. It certifies that a high-quality solution is
available to the specified shallow schedule, but it does not certify that the
variational dynamics will place substantial amplitude on that solution. The
remaining gap between structural opportunity and realized performance can
depend on interference, parameter optimization, probability concentration,
and finite-shot sampling.

Likewise, the observed performance of the objective-improving policy is used
here as a constructive demonstration of useful transition allocation, not as
evidence that objective-informed scheduling is universally optimal. The more
important result is policy independent: among the accessibility descriptors
tested here, identifying \emph{what objective quality is reachable} provides a
substantially sharper description of shallow-QAOA behavior than counting
mixer interactions or pair coverage alone.

A useful mechanism should survive changes in the initial state and problem
instance. We next test the same interpretation across alternative
feasible warm starts, nested portfolio sizes, and a non-financial synthetic
block-QUBO family.

\section{Robustness and Generalization}
\label{sec:robustness}

The canonical experiment establishes that mixer schedules with the same
executed interaction budget can expose finite-depth accessible regions of
substantially different objective quality. A useful accessibility framework,
however, should not depend on one specially chosen initial state, one
portfolio dimension, or the financial form of the objective. We test the conclusions of Sec.~\ref{sec:results} along three progressively
broader axes: the feasible initial state, the size of the constrained
portfolio instance, and finally the objective family itself.

These tests are structured to change one aspect of the problem
at a time. The warm-start study keeps the $n=20$ portfolio instance and
transaction-cost objective fixed while varying $x_0$. The nested study keeps
the portfolio construction but changes the number and arrangement of the
sector blocks. The synthetic study retains the block-cardinality feasible
structure and mixer construction while replacing the financial objective by
independently generated quadratic binary objectives.
Table~\ref{tab:robustness_design} summarizes this hierarchy.

\begin{table*}[t]
\caption{
Robustness hierarchy used to test the finite-depth accessibility mechanism.
The QAOA counts refer to completed favorable-tail CVaR runs. Structural
accessibility quantities are evaluated independently of the variational
optimization.
}
\label{tab:robustness_design}
\small
\begin{ruledtabular}
\begin{tabular}{lcccc}
Study
&
Sizes
&
QAOA rows
&
Varied quantity
&
Retained structure
\\
\hline
Warm-start robustness
&
$20$
&
83
&
initial state $x_0$
&
objective, $r$, block constraints
\\
Nested portfolio
&
$14,16,20$
&
40
&
size and block structure
&
portfolio setup, mixer rules
\\
Synthetic block-QUBO
&
$12,16,20$
&
68
&
objective coefficients
&
block constraints, mixer rules, $p=3$
\\
\end{tabular}
\end{ruledtabular}
\end{table*}

\subsection{Dependence on the Feasible Initial State}
\label{subsec:warmstart_robustness}

The initial state is known to be an important component of variational
quantum optimization. Classical warm starts can improve low-depth QAOA
performance \cite{Egger2021}, while for constrained QAOA the
alignment between the initial state and the mixer Hamiltonian can materially
change optimization behavior \cite{He2023}. Recent Lie-algebraic
analyses of XY-mixer QAOA further emphasize the interaction between mixer
structure, initialization, and trainability \cite{kordonowy2026lie}. Our
question is different but complementary: how strongly does the
\emph{finite-depth accessible region itself} depend on the feasible state
from which the executed schedule begins?

We evaluate six feasible initial states on the same $n=20$ portfolio
instance:
\begin{enumerate}
    \item the canonical first-two-per-sector incumbent;
    \item a return-ranked feasible state;
    \item a greedily constructed objective-based state; and
    \item three independently generated random feasible states.
\end{enumerate}
The transaction-cost reference portfolio $r$ is held fixed at the canonical
incumbent throughout. Consequently, replacing $x_0$ changes the state from
which mixer accessibility is evaluated but does not change the portfolio
objective $f(x)$. This separation is essential: otherwise a change in QAOA
initialization would simultaneously change the optimization landscape against
which the schedules are being compared.

The accessible set is expected to depend on $x_0$ because the set of active
exchanges in a mixer round,
\begin{equation*}
    A_\ell(x)
    =
    \left\{
        (i,j)\in M_\ell:
        x_i\neq x_j
    \right\},
\end{equation*}
is state dependent. Changing the initial occupation pattern can therefore
alter both the first-round transitions and every subsequent set
$R_\ell(M,x_0)$ generated by the recursion of
Sec.~\ref{subsec:round_accessibility}. The warm-start experiment confirms
that this dependence is substantial in practice.

It is particularly visible for the objective-improving policy. Because this
policy retains only initially positive-gain exchanges and constructs each
round from the same static ranked candidate list, its ability to realize a
requested budget can depend strongly on the initial state. For some initial
states or problem instances, one or more layer quotas may exceed the number
of positive-gain pairs that can be selected while maintaining
vertex-disjointness within that round. Since the OI ranking is static and
only positive-gain candidates are retained, this can yield
\begin{equation*}
    B_{\rm eff}<B.
\end{equation*}

For example, the greedy-objective start admits only three executed
objective-improving interactions under the implemented three-layer
construction, whereas the return-ranked start supports fifteen at the
high-budget setting. One of the random feasible starts supports nine.
Thus a requested budget such as $B=16$ is not always achievable under the
positive-gain OI selection rule from every $x_0$.

We retain this behavior and do not pad the schedule with zero- or
negative-gain interactions. A configuration for which the requested budget
cannot be executed is labeled \emph{capacity limited} and is excluded from a
nominally resource-matched comparison unless an explicit capacity-matched
comparison is made at the realized budget,
\begin{equation}
    B_{\rm eff}
    =
    \sum_{\ell=1}^{p}|M_\ell|
    <
    B .
    \label{eq:capacity_limited_budget}
\end{equation}

This observation is itself informative. Finite-depth accessibility is not a
property of a mixer topology in isolation. It depends jointly on
\begin{equation*}
    \boxed{
        (x_0,M_1,\ldots,M_p)
    }
\end{equation*}
and, for an objective-informed schedule, on the objective landscape used to
select those transitions.

Across the 83 completed warm-start QAOA configurations, the mean realized
effective score is $0.7806$. This aggregate value is descriptive; it is not
used to compare mixer policies because it combines different initial states,
budgets, and schedules. More importantly for the present argument,
the warm-start rows contribute to the dependence-aware cross-domain analysis
of Sec.~\ref{subsec:cluster_association}, in which $Q_R$ remains the strongest
accessibility descriptor tested after controlling for problem size, executed
XY budget, and problem domain.

The warm-start study does not support the stronger claim that a fixed schedule
policy is insensitive to initialization. Its more useful message is
mechanistic: \emph{changing the feasible initial state changes finite-depth
structural opportunity, and $R_p(M,x_0)$ captures that dependence while
keeping it separate from the fixed objective reference.}

\subsection{Nested Portfolio Instances}
\label{subsec:nested_robustness}

We next ask whether the accessibility picture is specific to the canonical
20-variable instance. Starting from the same screened portfolio universe, we
construct deterministic nested instances with
\begin{equation*}
    n\in\{14,16,20\}.
\end{equation*}
The corresponding block structures and quotas are
\begin{align}
    n=14:&
    \qquad
    (|\mathcal{B}_g|)
    =
    (4,4,4,2),
    &
    (k_g)
    &=
    (2,2,2,1),
    \nonumber\\
    n=16:&
    \qquad
    (|\mathcal{B}_g|)
    =
    (4,4,4,4),
    &
    (k_g)
    &=
    (2,2,2,2),
    \nonumber\\
    n=20:&
    \qquad
    (|\mathcal{B}_g|)
    =
    (4,4,4,4,4),
    &
    (k_g)
    &=
    (2,2,2,2,2).
    \label{eq:nested_block_structures}
\end{align}

Their feasible-set sizes are therefore
\begin{equation}
    |\mathcal F_{14}|=432,
    \qquad
    |\mathcal F_{16}|=1296,
    \qquad
    |\mathcal F_{20}|=7776.
    \label{eq:nested_feasible_sizes}
\end{equation}
For each nested universe, the portfolio return and covariance quantities are
recomputed on the corresponding asset subset instead of being inherited by
simply truncating the final Ising Hamiltonian.

The same fixed, rotating-disjoint, random-disjoint, and
objective-improving schedule constructions are evaluated at the matched
budgets used in the canonical comparison. This produces 40 completed QAOA
configurations in the nested robustness study, in addition to the exact
accessibility analysis performed for the corresponding schedules.

These three values of $n$ are not used to infer asymptotic scaling. The narrower
question is whether the finite-depth distinction between structural exposure
and objective-relevant accessibility survives a change in the feasible
manifold. It does. Changing $n$ changes both the number of feasible configurations
and the objective landscape, and consequently changes the absolute values of
$|R_p|$, $Q_R$, and $S_{\rm eff}$. Nevertheless, the nested instances
continue to exhibit substantial schedule dependence at fixed depth and
matched executed interaction budgets. In particular, increasing structural
coverage does not remove the need to ask which objective values occur inside
the region made accessible by the executed schedule.

The nested study also exposes a second form of finite-resource limitation.
For the $n=14$ instance at requested budget $B=16$, the
objective-improving construction cannot fill every requested layer quota
under its positive-gain, vertex-disjoint selection rule. The realized total is
\begin{equation*}
    B_{\rm eff}=15 < 16.
\end{equation*}
We therefore do not compare this row to genuinely $B=16$ schedules as though
the executed resources were identical. The requested-budget case is retained
as a capacity-limited diagnostic, while the realized $B_{\rm eff}=15$
condition can be examined separately.

Across all 40 completed nested QAOA configurations, the mean effective score
is $0.7126$. As in the warm-start study, this aggregate value is not used to
argue that one mixer policy dominates across sizes. The more relevant result
is that the schedule-specific accessibility construction remains
well-defined and discriminative as the feasible manifold changes, while
resource limitations are exposed explicitly and are not hidden in nominal
budget labels.

\subsection{Generalization Beyond the Portfolio Objective}
\label{subsec:synthetic_generalization}

The preceding tests still share the financial return--risk objective.
Consequently, a remaining concern is that useful accessibility might be a
special feature of the covariance structure or transaction-cost model in
Eq.~\eqref{eq:portfolio_objective}. To separate the accessibility mechanism
from that application, we construct a family of synthetic
block-constrained quadratic binary optimization problems.

Quadratic binary and equivalent Ising formulations provide a broad language
for combinatorial optimization \cite{Lucas2014}. For the synthetic
instances we maximize, using the symmetric-matrix convention of the
implementation,
\begin{equation}
    f_{\rm syn}(x)
    =
    h^\top x
    +
    x^\top Jx,
    \label{eq:synthetic_objective}
\end{equation}
subject to the same type of block-cardinality constraints used throughout
this work. The coefficients are generated independently as
\begin{equation*}
    h_i\sim{\rm Uniform}(-1,1),
\end{equation*}
and, for each unordered pair $i<j$,
\begin{equation}
    J_{ij}=J_{ji}
    \sim
    {\rm Uniform}(-1,1),
    \qquad
    J_{ii}=0.
    \label{eq:synthetic_coefficients}
\end{equation}
No coefficient rescaling is applied; comparisons use objective gaps
normalized separately within each instance.

We generate instances at
\begin{equation*}
    n\in\{12,16,20\},
\end{equation*}
using size-four blocks with quota two in every block. Thus
\begin{equation*}
    |\mathcal F_{12}|=216,
    \qquad
    |\mathcal F_{16}|=1296,
    \qquad
    |\mathcal F_{20}|=7776.
\end{equation*}
Three independent problem seeds, $101$, $202$, and $303$, are used for the
analytical accessibility study. QAOA confirmation is performed on the first
two seeds, giving 68 completed QAOA configurations after retaining the same
resource-matching and capacity rules used for the portfolio experiments.

This experiment changes the objective landscape substantially while
preserving the central combinatorial structure required by the theory: a
feasible computational-basis initial state, fixed block cardinalities,
diagonal cost evolution, and executed block-local XY mixer rounds. The same
quantities
\begin{equation*}
    R_p,
    \qquad
    \frac{|R_p|}{|\mathcal F|},
    \qquad
    f_R^\star,
    \qquad
    Q_R
\end{equation*}
can therefore be evaluated without reference to returns, covariance
matrices, sectors, or transaction costs.

The 68 synthetic QAOA configurations have mean realized effective score
$0.7135$. The aggregate score itself is not the main result. The synthetic rows enter the same harmonized accessibility analysis
as the portfolio experiments in
Sec.~\ref{subsec:mechanistic_association}. Despite the change in objective
family, the combined analysis yields
\begin{equation*}
    \rho_{\rm P}(Q_R,S_{\rm eff})=0.7003,
    \qquad
    \rho_{\rm S}(Q_R,S_{\rm eff})=0.8057,
\end{equation*}
and the partial association remains $0.6996$ after controlling for $n$,
executed XY budget, and domain. Because these controls explicitly distinguish
the synthetic and portfolio domains, the observed relationship cannot be
attributed simply to both families sharing the same raw objective scale.

The synthetic study provides evidence that the distinction between
\emph{how much is accessible} and \emph{what quality is accessible} is not
confined to the particular financial objective used in the primary
benchmark. It does not establish universality across arbitrary constrained
QUBOs, but it extends the empirical mechanism beyond the application from
which the study was motivated.

\subsection{Scope of the Robustness Evidence}
\label{subsec:robustness_scope}

The three studies move progressively farther from the canonical experiment:
from different feasible initial states for the same problem, to different
portfolio sizes, and finally to a different objective family.

The evidence supports three conclusions.

First, finite-depth accessibility is intrinsically initial-state dependent.
This is not a nuisance parameter to be removed from the framework:
$R_p(M,x_0)$ is intended to expose precisely this dependence.

Second, the accessibility analysis remains meaningful as the number and size
of feasible blocks change. The nested experiments also demonstrate why
actual executed interaction counts, not just requested budgets, are needed for
fair comparisons.

Third, replacing the financial objective with independently generated
quadratic objectives does not remove the empirical relevance of
objective-aware accessibility. Together with the controlled and
dependence-aware results of Sec.~\ref{sec:results}, this supports
finite-depth reachable quality as a useful mechanistic diagnostic for the
tested family of block-constrained shallow-QAOA circuits.

These conclusions remain narrower than a scaling or
universality claim. The numerical studies are restricted to $p=3$,
$n\leq20$, block-cardinality constraints, and the particular portfolio and
synthetic objective families described above. Exact enumeration of
$R_p$ and $\mathcal F$ is also practical only because these instances remain
moderate in size. Extending useful-accessibility diagnostics to regimes in
which exact enumeration is impossible will require estimators, bounds, or
sampling-based approximations.

The remaining practical question is different: whether schedules that expose
higher-quality finite-depth regions require substantially greater physical
circuit resources once hardware connectivity and routing are taken into
account. We address that question next.

\section{Hardware and Compilation Implications}
\label{sec:hardware}

The preceding results establish that mixer schedules with comparable logical
interaction budgets can expose finite-depth accessible regions of substantially
different objective quality. This raises a practical question that is distinct
from the variational optimization problem itself: does obtaining better
finite-depth accessibility require substantially greater physical circuit
resources once limited connectivity and routing are taken into account?

This distinction is important for QAOA because logical circuit structure does
not translate directly into hardware cost. Limited processor connectivity can
introduce additional routing operations, and the resulting overhead depends on
both the interaction graph of the problem Hamiltonian and the structure of the
variational ansatz \cite{Lotshaw2022,Ji2025}. The issue is especially relevant
for dense optimization Hamiltonians, for which even a shallow logical circuit
may require a large number of routed two-qubit interactions
\cite{Maciejewski2024}. Recent QAOA-specific compilation methods based on SWAP
networks and parity-based encodings further demonstrate that the physical cost
of nonlocal interaction graphs can depend strongly on the compilation strategy
\cite{MontanezBarrera2026}.

We do not propose a new compiler or report execution on a quantum processor.
The compilation study is a resource audit of how the objective quality exposed
by different feasible mixer schedules interacts with hardware connectivity. In particular, we separate the compiled cost of
the phase separator from that of the constraint-preserving mixer.

\subsection{Compilation Protocol}
\label{subsec:hardware_protocol}

We compile the portfolio circuits associated with the frozen
resource-controlled mixer schedules studied in the preceding sections.
The analysis includes the nested portfolio sizes
\begin{equation*}
    n\in\{14,16,20\},
\end{equation*}
and preserves the same executed mixer schedules used in the corresponding
finite-depth accessibility calculations. No variational reoptimization is
performed during the compilation study.

Three connectivity models are considered. The first is an all-to-all
connectivity model, which provides a logical reference in which every
two-qubit interaction can be implemented directly. The second is a
two-dimensional nearest-neighbor mesh, representing a regular sparse
connectivity constraint. The third is a lower-degree hardware-inspired sparse
graph used as an additional routing stress test. The latter two are synthetic
connectivity models, not coupling maps of specific quantum processors. The
results should therefore be interpreted as topology-dependent compilation
diagnostics, not as device benchmarks.

Circuits are transpiled using Qiskit 2.2.3 with optimization level 1. For the
connectivity-constrained cases, SABRE layout and routing are used with
transpiler seeds
\begin{equation*}
    s_{\rm tr}\in\{42,43,44\}.
\end{equation*}
The target basis is
\begin{equation*}
    \{\mathrm{RZ},\mathrm{SX},X,\mathrm{CX}\}.
\end{equation*}
Multiple transpiler seeds are retained because heuristic placement and routing
can otherwise make a single compiled circuit an unnecessarily fragile measure
of physical resource cost.

The compilation analysis is performed separately for three circuit
components:
\begin{enumerate}
    \item the phase-separation or cost circuit generated by $H_C$;
    \item the block-local XY mixer circuit generated by the executed schedule;
    \item the complete shallow-QAOA circuit containing both components.
\end{enumerate}

Let
\begin{equation*}
    N_{2q}^{(C)},\qquad
    N_{2q}^{(M)},\qquad
    N_{2q}^{(\mathrm{full})}
\end{equation*}
denote the resulting two-qubit gate counts for the separately transpiled cost,
mixer, and complete circuits, respectively. We use the diagnostic ratios
\begin{equation}
    \eta_C
    =
    \frac{N_{2q}^{(C)}}{N_{2q}^{(\mathrm{full})}},
    \qquad
    \eta_M
    =
    \frac{N_{2q}^{(M)}}{N_{2q}^{(\mathrm{full})}}.
    \label{eq:hardware_cost_fractions}
\end{equation}
Because the isolated components and the full circuit are transpiled
independently, these quantities are not intended as an additive decomposition:
in general,
\begin{equation*}
    \eta_C+\eta_M\neq 1.
\end{equation*}
Compiler optimizations, layout choices, and routing interactions can differ
between the isolated and complete circuits. The ratios instead identify which
logical component sets the dominant two-qubit resource scale.

This protocol does not include device noise, calibration data,
measurement error, or hardware execution. The resulting quantities measure
compilation overhead; they do not measure physical fidelity.

\subsection{Dense Phase Separation Dominates the Compiled Cost}
\label{subsec:cost_dominance}

The portfolio objective of
Eq.~\eqref{eq:portfolio_objective} contains the covariance term
$x^\top\Sigma x$. For the market-derived instances studied here, this produces
a dense set of pairwise Ising couplings in the phase-separation Hamiltonian,
\begin{equation*}
    H_C
    =
    \sum_i h_i Z_i
    +
    \sum_{i<j}J_{ij}Z_iZ_j
    +
    C.
\end{equation*}
The constraint-preserving mixer, in contrast, contains only the selected
block-local XY interactions allocated by the schedule. The two circuit
components therefore have very different interaction graphs before any
hardware topology is imposed.

The compilation results show that this asymmetry persists, and in some cases
becomes stronger, after routing. Table~\ref{tab:hardware_cost_fraction}
summarizes the cost-to-full two-qubit ratio for the $n=20$ portfolio circuits.

\begin{table}[t]
\caption{
Cost-to-full two-qubit compilation ratio for the $n=20$ portfolio study.
Cost-only and full circuits are transpiled independently, so the values are
diagnostic ratios and should not be read as additive gate fractions.
}
\label{tab:hardware_cost_fraction}
\begin{ruledtabular}
\begin{tabular}{lc}
Connectivity model
&
$\eta_C=N_{2q}^{(C)}/N_{2q}^{(\mathrm{full})}$
\\
\hline
All-to-all
&
$0.960$
\\
Two-dimensional mesh
&
$0.992$
\\
Sparse hardware-inspired
&
$0.959$
\\
\end{tabular}
\end{ruledtabular}
\end{table}

Even under all-to-all connectivity, the cost circuit already contributes a
two-qubit count comparable to that of the complete circuit,
\begin{equation*}
    \eta_C\simeq0.960.
\end{equation*}
This dominance is not solely a routing artifact: it originates from
the dense financial interaction graph itself.

The effect is most pronounced on the two-dimensional mesh, where the
cost-to-full ratio rises to approximately
\begin{equation*}
    \eta_C\simeq0.992.
\end{equation*}
Here many of the dense $ZZ$ interactions cannot be realized directly on
neighboring physical qubits, so the phase separator incurs substantial routing
overhead. The sparse hardware-inspired topology exhibits the same qualitative
behavior, with
\begin{equation*}
    \eta_C\simeq0.959.
\end{equation*}

The same cost-dominated structure is observed across the nested portfolio
sizes. On the sparse hardware-inspired topology, the corresponding
cost-to-full ratios are approximately
\begin{equation*}
    0.944,\qquad
    0.956,\qquad
    0.959
\end{equation*}
for $n=14$, $16$, and $20$, respectively. These values are specific to this
synthetic sparse topology and the compilation protocol described above; they
are not averages over all connectivity models. Within this controlled
compilation study, the dense phase separator sets the leading two-qubit
resource scale for the portfolio instances, while the feasibility-preserving
mixer remains secondary in the compiled count.

This observation is consistent with the broader hardware challenge posed by
dense Ising optimization. Compilation-aware implementations can substantially
alter the cost of such interaction structures
\cite{Maciejewski2024,Ji2025,MontanezBarrera2026}, but the present experiment
does not attempt to identify an optimal compilation strategy. Instead, it
establishes which component of our constrained-QAOA circuit would require the
greatest attention in a hardware-aware implementation.

\subsection{Accessibility Remains a Distinct Mixer Design Axis}
\label{subsec:hardware_accessibility}

Cost-layer dominance does not imply that the mixer is algorithmically
irrelevant. The phase separator and mixer play fundamentally different roles.
The phase separator assigns phases according to the objective landscape,
whereas the block-local XY schedule determines the feasible basis
configurations that can become structurally accessible from $x_0$ within the
available depth.

The resource-controlled experiments match policies using the
number of executed logical XY interactions,
\begin{equation*}
    B
    =
    \sum_{\ell=1}^{p}|M_\ell|.
\end{equation*}
Consequently, differences in $R_p(M,x_0)$ and $Q_R$ cannot be attributed
simply to one policy receiving a larger logical mixer budget. The compilation
study adds a second observation: after mapping those schedules to sparse
connectivity, the isolated mixer cost remains comparatively small relative to
the dense phase separator.

For the $n=20$ connectivity-constrained cases, the mean isolated
mixer-to-full ratios are approximately
\begin{equation*}
    \eta_M\simeq0.027\text{--}0.030,
\end{equation*}
for the sparse hardware-inspired and two-dimensional mesh topologies,
respectively. Individual compiled rows vary with mixer schedule and transpiler
seed, but remain at only a few percent of the corresponding full-circuit
two-qubit count.

\begin{figure*}[t]
    \centering
    \includegraphics[width=\textwidth]{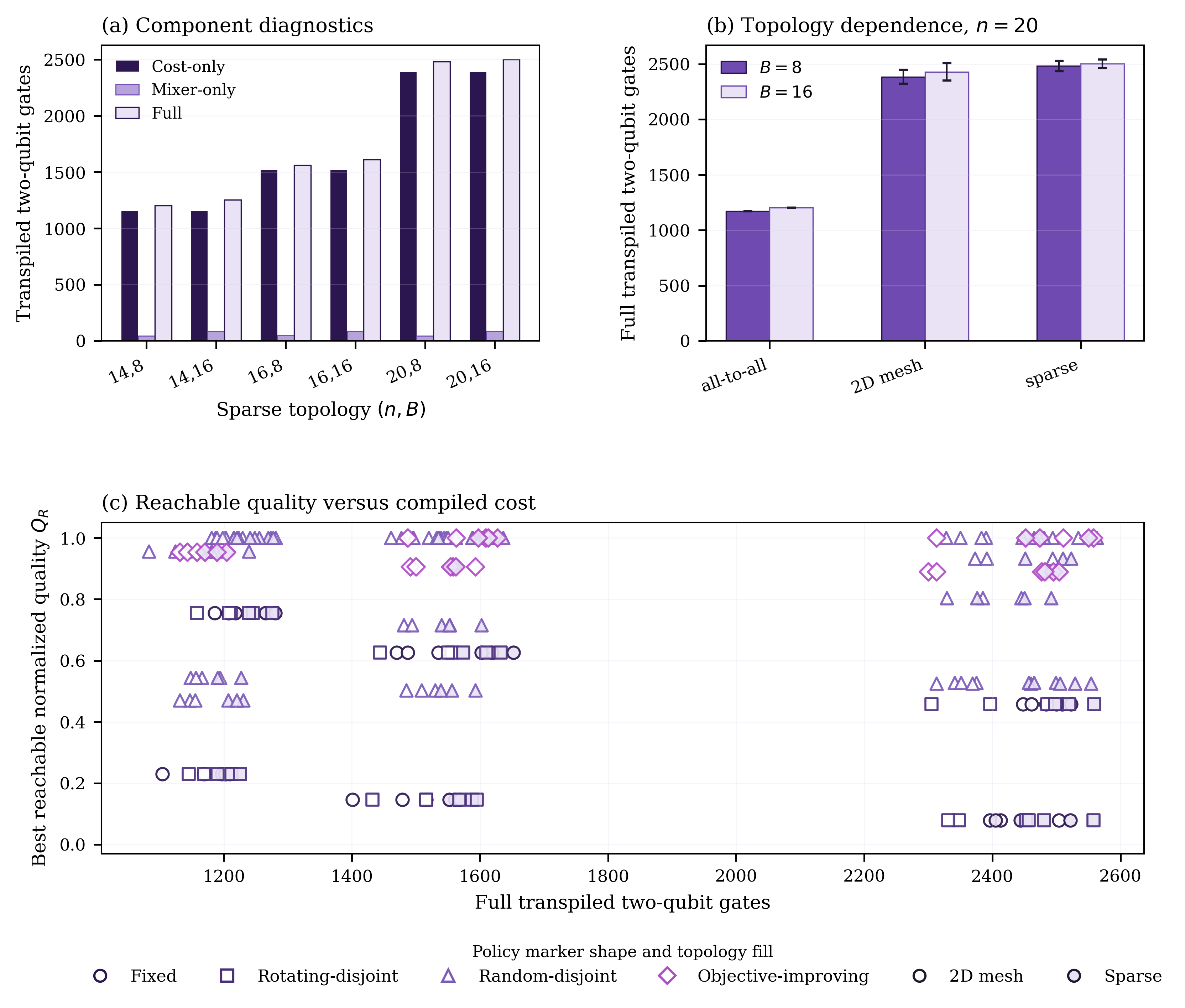}
    \caption{
    Compilation diagnostics for the nested portfolio circuits.
(a) Two-qubit gate counts for the separately transpiled cost, mixer,
and full circuits on the sparse hardware-inspired topology.
(b) Full-circuit two-qubit counts for $n=20$ across the three
connectivity models and the two logical XY budgets. Error bars show
the sample standard deviation across the retained compiled rows in
each topology--budget group.
(c) Best reachable normalized quality $Q_R$ versus full transpiled
two-qubit gate count for the connectivity-constrained cases; marker
shape identifies the mixer policy and marker fill distinguishes the
two-dimensional mesh from the sparse hardware-inspired topology.
    }
    \label{fig:hardware_compilation}
\end{figure*}

This creates an important separation between \emph{algorithmic value} and
\emph{dominant physical cost}. Mixer schedules can expose substantially
different objective quality at finite depth even when their contribution to
the total compiled two-qubit count is secondary. Figure~\ref{fig:hardware_compilation}
summarizes this separation. Panel (a) shows that the dense cost layer sets the
dominant two-qubit resource scale across the nested portfolio instances, while
panel (b) shows the additional dependence of the full compiled cost on
connectivity. Panel (c) further shows that comparable compiled two-qubit counts
can correspond to substantially different values of $Q_R$. Compiled circuit cost alone does not determine the objective
quality exposed by a finite-depth mixer schedule.
Comparable physical resource scale does not imply comparable finite-depth
accessible quality.

For mixer design, minimizing gate count alone is insufficient. A mixer with
very low physical cost can still be algorithmically poor if its finite-depth
accessible set excludes high-quality solutions. Conversely, allocating the same number of logical interactions to
more objective-relevant feasible exchanges can raise $Q_R$ without requiring a
commensurate increase in the total compiled circuit cost.

This interpretation is policy independent. The
objective-improving construction provides one useful demonstration of such
transition allocation, but the compilation results do not establish that this
policy is universally optimal or hardware minimal. The result is instead a
separation between \emph{useful accessibility} and \emph{compiled resource
cost}: both should be evaluated, and neither substitutes for the other.

A useful hardware-facing view is the pair
\begin{equation}
    \left(
        N_{2q}^{(\mathrm{full})},
        Q_R
    \right),
    \label{eq:hardware_quality_pair}
\end{equation}
or, when a scalar resource-efficiency diagnostic is useful,
\begin{equation}
    \mathcal{E}_R
    =
    \frac{Q_R}{N_{2q}^{(\mathrm{full})}}.
    \label{eq:reachable_quality_efficiency}
\end{equation}
We use Eq.~\eqref{eq:reachable_quality_efficiency} only as an interpretive
diagnostic; it is not proposed as a universal objective for compiler or mixer
optimization.

\subsection{Implications and Scope}
\label{subsec:hardware_scope}

The compilation study suggests a two-axis view of hardware-aware constrained
QAOA. The first axis concerns the cost Hamiltonian: dense objective couplings
must be placed, ordered, routed, or compressed efficiently. This problem is
closely related to existing work on algorithm-aware qubit mapping and
specialized compilation of dense and partially connected QAOA interaction
graphs \cite{Ji2025,Maciejewski2024,MontanezBarrera2026}. The second axis
concerns the mixer: given a limited transition budget, its interactions should
be chosen so that the finite-depth accessible region contains useful
solutions.

These two objectives should not be conflated. Reducing mixer depth alone does
not solve the routing problem created by a dense phase separator. Likewise,
optimizing the dense cost circuit does not determine whether the shallow mixer
schedule exposes high-quality feasible configurations. For the portfolio
instances studied here, the former dominates the physical two-qubit count
while the latter remains consequential for finite-depth structural
opportunity.

Several limitations define the boundary of this result. First, the analysis is
compilation-only; no claim is made about noisy-device success probability,
fidelity, or quantum advantage. Second, the mesh and sparse connectivity
models are synthetic abstractions and should not be interpreted as
representations of a particular processor. Third, the numerical gate counts
depend on the selected basis, transpiler configuration, optimization level,
and routing heuristic. More specialized compilation methods can produce
different absolute resource requirements
\cite{Ji2025,MontanezBarrera2026}. Finally, the observed dominance of the
phase separator follows from the dense covariance structure of the portfolio
objective. A sparse or geometrically local objective can exhibit a very
different cost--mixer balance.

Subject to these limitations, the practical conclusion is that, for the
portfolio circuits considered here, choosing a mixer schedule with better
finite-depth objective accessibility does not introduce a comparable increase
in the dominant compiled two-qubit resource. Instead, dense phase-separator
compilation sets the principal hardware cost, while mixer scheduling remains a
secondary but independent algorithmic design axis. Hardware-aware constrained
QAOA should therefore consider \emph{where useful feasible states become
accessible} and \emph{how the objective interactions are physically compiled}
as related but distinct design problems.

This separation also clarifies the scope of the present work. Finite-depth
accessibility characterizes structural opportunity before noise, while
compilation determines one important part of the cost of realizing that
opportunity on restricted connectivity. Their combination motivates the
broader design principles and limitations discussed in the following section.

\section{Discussion}
\label{sec:discussion}

The results of the preceding sections support a view of constrained QAOA in
which feasibility preservation is only the first of several structural
questions. A mixer may preserve the feasible manifold exactly, and the
transition graph generated by its available interactions may even be globally
connected, while a shallow circuit initialized at a particular feasible basis
state still accesses only a schedule-dependent portion of that manifold. The
central object used in this work,
\begin{equation*}
    R_p(M,x_0),
\end{equation*}
makes this finite-depth restriction explicit. It depends jointly on the
initial feasible configuration $x_0$, the ordered sequence of executed mixer
rounds $M=(M_1,\ldots,M_p)$, and the available depth $p$.

This perspective complements the established constrained-mixer literature. The quantum alternating operator framework
provides a general construction for optimization dynamics restricted to
feasible sectors \cite{hadfield2019alternating}, while subsequent work has
developed and analyzed constraint-preserving mixers and the transitions they
induce \cite{fuchs2022constrained}. More recent studies have emphasized
transition-graph structure, mixer topology, initialization, and the dynamical
properties of constrained ans\"atze
\cite{gupta2026rostering,kordonowy2026lie}. Our focus is the gap between global transition structure and the
configurations exposed by the \emph{particular finite sequence of mixer
interactions actually executed} by a shallow circuit from a specified initial
state.

The resulting progression runs from feasibility, to global connectivity, to
finite-depth accessibility, to objective-relevant accessibility, and finally
to realized QAOA performance. These are successive questions rather than
equivalent descriptions; each adds information that is absent from the
preceding level.

\subsection{Structural Opportunity Versus Variational Attainment}
\label{subsec:discussion_opportunity}

The support-containment result provides the clearest interpretation of
finite-depth accessibility. For the fixed schedules considered here,
\begin{equation*}
    \operatorname{supp}
    \left(
        P_{\boldsymbol{\beta},\boldsymbol{\gamma}}
    \right)
    \subseteq
    R_p(M,x_0)
\end{equation*}
for every choice of variational parameters. Consequently, if a feasible
configuration lies outside $R_p(M,x_0)$, no adjustment of the continuous
QAOA angles can assign measurement probability to that configuration while
the circuit structure is held fixed. The limitation is structural and
parameter independent.

The containment can be strict. Membership in $R_p$ does not imply nonzero
probability for every parameter choice, nor does it imply that variational
optimization will place appreciable probability on the corresponding basis
state. Special angles, destructive interference, and correlations among
multiple quantum paths can suppress states that are structurally accessible.
This distinction is closely related to recent work separating geometric
transport structure from coherent interference in constrained quantum
optimization \cite{onah2026geometry}. In the terminology used here,
$R_p(M,x_0)$ specifies where amplitude is structurally permitted to appear;
it does not determine how that amplitude is distributed.

The same distinction applies to
\begin{equation*}
    f_R^\star
    =
    \max_{x\in R_p(M,x_0)} f(x)
\end{equation*}
and to its normalized form $Q_R$. These quantities characterize the best
objective quality structurally available within the accessible set. They do
not specify the objective value that the optimized quantum state will
actually sample. The fundamental ceiling is instead
\begin{equation}
    f_{\rm best}
    \leq
    f_R^\star ,
    \label{eq:discussion_raw_ceiling}
\end{equation}
for any best-of-samples statistic obtained from the fixed ideal circuit.

The two normalized quantities serve different purposes: $Q_R$ measures
structural opportunity from the actual initial state, while $S_{\rm eff}$
reports realized best-sample performance using the experiment's fixed
reporting reference.
When the two quantities are constructed using the same normalization
reference, Eq.~\eqref{eq:discussion_raw_ceiling} induces the corresponding
normalized ceiling. In the warm-start robustness study, however,
$Q_R$ is normalized relative to the actual initial state while the reported
$S_{\rm eff}$ uses the fixed experimental reference floor. We therefore do
not interpret $Q_R$ as a universal rowwise upper bound on the reported
$S_{\rm eff}$; the invariant structural statement is
Eq.~\eqref{eq:discussion_raw_ceiling}.

Several mechanisms intervene between structural opportunity and realized
attainment. Even when $Q_R=1$, so that a globally optimal feasible
configuration belongs to $R_p$, successful sampling may still be limited by
interference, parameter optimization, probability concentration, and
finite-shot fluctuations. On a physical processor, control error,
decoherence, and measurement noise introduce additional mechanisms.
Conversely, poor reachable quality imposes a more fundamental restriction:
if all high-quality configurations are excluded from $R_p$, modifying the
continuous variational parameters alone cannot make those configurations
appear.

The CVaR experiments provide empirical support for the usefulness
of this distinction. Across the portfolio and synthetic experiments, $Q_R$
shows the strongest association with realized shallow-QAOA performance among
the structural accessibility diagnostics considered here, with pooled Pearson
and Spearman associations of $0.7003$ and $0.8057$, respectively, and a
controlled association of $0.6996$. The dependence-aware analysis gives a
cluster-robust coefficient of $0.5893$ with $p=0.0031$. These results are
observational: they support $Q_R$ as an informative structural diagnostic in
the tested settings, not as a causal variable or a universal predictor of
QAOA performance.

The realized-performance layer also depends on experimental choices that do
not alter the structural quantities themselves. The variational objective,
optimizer budget, parameter initialization, shot count, and sampling protocol
can all affect $S_{\rm eff}$ while leaving $R_p$, $f_R^\star$, and $Q_R$
unchanged for a fixed problem, initial state, and executed mixer schedule.

\subsection{Accessible Quantity Is Not Accessible Quality}
\label{subsec:discussion_quantity_quality}

A central consequence of the framework is that the size of the accessible
set is not, by itself, an adequate description of mixer quality. Pair
coverage and reachable-space fraction characterize how broadly a schedule
can expose the feasible manifold, but they do not determine which objective
values occur within the exposed region. Two schedules may therefore satisfy
\begin{equation*}
    |R_p^{(1)}|
    <
    |R_p^{(2)}|
\end{equation*}
while simultaneously satisfying
\begin{equation*}
    f_{R,1}^\star
    >
    f_{R,2}^\star.
\end{equation*}

The empirical results do not imply that accessible volume is
irrelevant. Reachable fraction remains positively associated with realized
performance, including under the dependence-aware analysis. The more precise
conclusion is that accessible volume and accessible objective quality answer
different questions, and that the latter carries substantially stronger
information about realized shallow-QAOA performance in the experiments
studied here.

Connectivity remains important because disconnected transition structures
can permanently exclude feasible solutions. At finite depth, however, global
connectivity alone is insufficient. A globally connected transition graph can
still allocate the available interaction budget in directions that fail to
expose high-quality configurations from the chosen $x_0$. At finite depth, both the amount of feasible space exposed and the objective
quality inside that exposed region matter.

The separation becomes especially relevant when mixer interactions are
allocated selectively. Adaptive and selective mixer-allocation strategies
have already shown that applying every available mixer interaction at every
layer is not always the most resource-efficient choice
\cite{ni2026adaptive}. The objective-improving schedule used here is related
in motivation but is not adaptive during circuit execution. Candidate
single-swap gains are evaluated once from $x_0$, the ranking is fixed before
the circuit is executed, and the same static candidate ordering is used to
construct the scheduled rounds.

Its role is consequently narrower than that of a general adaptive mixer
method. OI provides a controlled intervention that changes the objective
content of $R_p$ while the requested interaction budget is held fixed whenever
the positive-gain disjoint capacity permits. It is not proposed as a
universally optimal mixer, and objective-informed interaction selection is
not itself the principal novelty of the present work.

The results suggest a practical design heuristic: allocate finite mixer resources with useful accessibility in mind instead of relying on connectivity alone. This is not an optimality theorem. Which transitions are useful can depend on the objective, initial state, available depth, and constraint structure, and obtaining useful transition scores may itself require nontrivial classical computation.

\subsection{Initial-State and Problem Dependence}
\label{subsec:discussion_dependence}

Finite-depth accessibility is inherently relative to an initial feasible
configuration. The notation $R_p(M,x_0)$ is intended to expose this
dependence explicitly. Changing $x_0$ can alter which scheduled mixer edges
are active in the first round, which configurations become available in later
rounds, and the objective values contained in the resulting accessible set.
This is consistent with prior observations that initialization and mixer
structure can materially affect constrained-QAOA dynamics
\cite{kordonowy2026lie}.

The warm-start experiments serve a different purpose from
identifying a universally preferred initialization. They test whether the
finite-depth accessibility picture survives movement of the starting point
within the same feasible manifold. Importantly, the QAOA initial state and
the transaction-cost reference are separate objects in the portfolio
experiments. Varying $x_0$ changes the origin of the accessibility calculation
without simultaneously redefining the objective through a new turnover
reference.

The resulting capacity-limited OI cases further illustrate why accessibility
must be indexed by the actual executed schedule rather than by a nominal
policy label alone. Because the OI candidate ranking is static and contains
only positive-gain exchanges, some initial states do not supply enough
vertex-disjoint positive-gain pairs to fill every requested layer quota. In
such cases,
\begin{equation*}
    B_{\rm eff}<B,
\end{equation*}
and the comparison is no longer resource matched at the requested budget.
This behavior is not a failure of the accessibility definition; it is
precisely the kind of schedule- and initial-state-specific structural
limitation that the framework is intended to expose.

Problem dependence is equally important. The portfolio benchmark combines a
dense covariance-derived objective with economically interpretable
block-cardinality constraints, whereas the synthetic block-QUBO experiments
replace the financial objective while retaining the same broad constraint
structure. Agreement across these settings provides evidence that the
quantity-versus-quality distinction is not an artifact of one covariance
matrix or one portfolio instance. It does not establish universality across
arbitrary constrained QUBOs, mixer Hamiltonians, constraint classes, or
circuit depths.

The theoretical and empirical claims should be separated. Support
containment follows from the structure of the executed constraint-preserving
mixer rounds and diagonal cost evolution and does not require a financial
objective. The empirical usefulness of a particular schedule constructor, or
the strength of the observed relationship between $Q_R$ and QAOA performance,
can depend strongly on the objective landscape and should not be promoted to
a problem-independent theorem.

\subsection{Scaling Beyond Exact Enumeration}
\label{subsec:discussion_scaling}

The exact accessibility analysis used in this work is intentionally
diagnostic. For block-cardinality constraints,
\begin{equation*}
    |\mathcal F|
    =
    \prod_g
    \binom{|\mathcal{B}_g|}{k_g},
\end{equation*}
and both the feasible manifold and the finite-depth accessible set can grow
combinatorially with problem size. Exact enumeration of $\mathcal F$ and
$R_p$ is therefore practical only for moderate instances.

This computational limitation does not remove the conceptual role of
finite-depth accessibility, but it changes how the associated quantities
would need to be estimated at larger scales. Partial exploration of $R_p$
can produce constructive lower bounds on $f_R^\star$, while certified upper
bounds could combine restricted reachability information with relaxations of
the objective over unexplored configurations. Layerwise algorithms may also
exploit block structure and the disjoint-round construction rather than
materializing the complete feasible manifold.

A useful large-scale extension would therefore replace exact $Q_R$ by
computable bounds,
\begin{equation*}
    Q_R^{\rm lower}
    \leq
    Q_R
    \leq
    Q_R^{\rm upper},
\end{equation*}
with costs that remain manageable when complete state enumeration is
impossible. Such bounds could preserve one of the principal diagnostic uses
of $Q_R$: identifying shallow schedules whose structural opportunity is
provably poor, without requiring the exact accessible set.

Sampling-based approximations provide another possibility. One may estimate
the distribution of objective values contained in partially explored
accessible regions or estimate whether high-quality regions are encountered
with increasing search effort. Such estimates would, however, need to be
distinguished carefully from certified values of $f_R^\star$ or $Q_R$.

The schedule-construction problem introduces a related scaling challenge.
Exact evaluation of every legal objective change is inexpensive for the
instances studied here but may become costly for larger or more complicated
objectives. Approximate local scores, classical relaxations, learned
transition priorities, or hardware-informed screening could therefore be
used to propose candidate schedules, while accessibility diagnostics provide
a separate structural evaluation of the schedules that result.

\subsection{Algorithmic and Hardware Design Should Remain Distinct}
\label{subsec:discussion_algorithm_hardware}

Section~\ref{sec:hardware} adds a practical qualification to the accessibility
picture. A mixer interaction can be algorithmically valuable without being
the dominant source of compiled circuit cost. For the dense portfolio
objectives studied here, the covariance-derived phase separator sets the
leading two-qubit resource scale, while the block-local mixer contributes
only a small fraction of the compiled full-circuit cost.

This observation is specific to the interaction structure examined here and
should not be generalized to all constrained-QAOA problems. Sparse objective
Hamiltonians, geometrically local cost functions, alternative encodings, or
different mixer constructions can yield a substantially different balance.
The more general point is that the two design roles answer different
questions. Cost-layer design asks how objective interactions can be implemented
efficiently; mixer design asks which useful feasible configurations become
accessible. Optimizing one axis does not automatically solve the other. Reducing the logical mixer
gate count does not eliminate routing overhead generated by a dense phase
separator, while compiling the phase separator efficiently does not ensure
that the finite-depth mixer schedule exposes high-quality feasible states.
A hardware-aware constrained-QAOA workflow should therefore treat these
questions as coupled but separately measurable design problems.

The present compilation study is only an initial resource analysis in that
direction. It uses synthetic connectivity graphs and ideal transpilation,
without device-specific calibration data or noisy hardware execution.
Finite-depth accessibility is likewise an ideal structural property: it
characterizes which computational-basis states can be reached by the circuit
architecture before physical noise is introduced. Extending the framework
to noisy execution would require distinguishing structural inaccessibility
from configurations that are structurally available but practically
suppressed by decoherence, control errors, routing overhead, or measurement
noise.

\subsection{Scope of the Contribution}
\label{subsec:discussion_scope}

The contribution of this work is a finite-depth diagnostic framework, not a
universal mixer prescription. We do not claim that
objective-improving schedules are optimal for constrained QAOA, that $Q_R$
fully determines variational performance, or that finite-depth accessibility
captures every feature relevant to quantum optimization. In particular, the
framework does not by itself characterize trainability, coherent
interference, parameter concentration, optimizer robustness, noise
sensitivity, or sampling complexity.

The framework isolates a structural question that precedes those effects. Before
asking whether an optimizer can locate favorable parameters, whether
interference can concentrate amplitude on desirable states, or whether a
physical processor can execute the resulting circuit accurately, one may
first ask whether the fixed shallow circuit can structurally access the
desired feasible configurations at all. The quantity $f_R^\star$ then provides a parameter-independent ceiling on
the best computational-basis objective obtainable from that fixed ideal circuit.

Finite-depth accessibility complements other levels of
constrained-QAOA analysis. Feasible-subspace constructions determine where
the dynamics are permitted to remain
\cite{hadfield2019alternating,fuchs2022constrained}; transition-graph and
Lie-algebraic analyses characterize connectivity and dynamical structure
\cite{gupta2026rostering,kordonowy2026lie}; interference-based analyses ask
how amplitudes combine along structurally available paths
\cite{onah2026geometry}; and compilation determines part of the physical cost
of realizing those paths on restricted hardware. Finite-depth accessibility
occupies an intermediate level: it asks which feasible computational-basis
states are structurally available to the \emph{actual resource-bounded
circuit}, and whether those states contain solutions of useful objective
quality.

The conclusion is narrower than a claim of quantum advantage but broader
than the portfolio application: preserving feasibility is necessary, while
useful shallow optimization also depends on which feasible solutions become
accessible at the available depth.

The implication for constrained mixer design is schedule-agnostic: no
particular schedule is preferred universally. Feasibility preservation and
global connectivity do not exhaust the design problem at finite depth. Under a limited interaction budget, the objective content of
the region exposed by the executed schedule is an additional quantity that
can be analyzed, compared, and ultimately optimized.

\section{Conclusion}
\label{sec:conclusion}

This work examined constrained QAOA from the perspective of the circuit that
is actually executed: a fixed feasible initial state, a finite sequence of
constraint-preserving mixer rounds, and a shallow depth. Within this setting,
we defined the finite-depth accessible set
\begin{equation*}
    R_p(M,x_0),
\end{equation*}
and established the structural containment
\begin{equation}
    \operatorname{supp}
    \left(
        P_{\boldsymbol{\beta},\boldsymbol{\gamma}}
    \right)
    \subseteq
    R_p(M,x_0)
    \subseteq
    \mathcal F.
    \label{eq:conclusion_support_containment}
\end{equation}
This separates the full feasible manifold from the configurations made
structurally accessible by the executed shallow schedule and from the subset
that ultimately receives nonzero probability. The corresponding best
reachable objective,
\begin{equation*}
    f_R^\star
    =
    \max_{x\in R_p(M,x_0)} f(x),
\end{equation*}
provides a parameter-independent ceiling on the best computational-basis
objective obtainable from the fixed ideal circuit. Its normalized form,
$Q_R$, characterizes the objective quality contained within the finite-depth
accessible region.

The resource-controlled experiments show why this distinction matters.
Schedules with comparable executed XY-interaction budgets can expose
substantially different accessible regions, and a larger accessible volume does not guarantee better objective values. Across the favorable-tail
CVaR portfolio and synthetic experiments, $Q_R$ exhibits the strongest
association with realized shallow-QAOA performance among the structural
diagnostics considered here. Reachable-space fraction also carries useful
information, while pair coverage becomes substantially weaker after
controlling for problem size, executed interaction budget, problem domain,
and dependence among repeated observations from the same instances. These
results support objective-relevant accessibility as a useful structural
diagnostic for the tested shallow circuits, without implying that $Q_R$
determines variational performance or that any particular mixer-allocation
policy is universally optimal.

The same picture persists under progressively broader changes to the
experimental setting. Varying the feasible initial state changes
$R_p(M,x_0)$ and can alter whether an objective-informed schedule has
sufficient positive-gain, vertex-disjoint capacity to realize a requested
interaction budget. Nested portfolio instances show that the distinction
between accessible volume and accessible quality remains visible as the
feasible manifold changes, while synthetic block-constrained quadratic
objectives extend the observation beyond the covariance-based financial
objective. The compilation study provides a complementary practical result:
for the dense portfolio Hamiltonians considered here, the phase separator
sets the dominant compiled two-qubit resource scale, while alternative
resource-controlled mixer schedules can still produce substantially different
finite-depth objective accessibility. Algorithmic accessibility and compiled
hardware cost are related but distinct design axes.

The resulting view of constrained QAOA is more specific than asking whether a
mixer preserves feasibility or whether its global transition graph is
connected. The structural relation is
\begin{equation}
\boxed{
    \operatorname{supp}
    \left(
        P_{\boldsymbol{\beta},\boldsymbol{\gamma}}
    \right)
    \subseteq
    R_p(M,x_0)
    \subseteq
    \mathcal F
}
\label{eq:conclusion_hierarchy}
\end{equation}
Global connectivity describes what
the mixer structure can connect in principle, whereas $R_p(M,x_0)$ describes
what the particular resource-bounded schedule can expose from its chosen
initial state at the available depth. The objective values inside that region
then determine whether the resulting structural opportunity is useful for the
optimization problem.

For shallow constrained QAOA, feasibility preservation and global connectivity
leave an additional design question. Under a limited interaction budget, which
feasible configurations does the actual finite-depth circuit make structurally
available, and do those configurations contain solutions of sufficient
objective quality to be worth reaching?

\bibliographystyle{apsrev4-2}
\bibliography{apssamp}

\end{document}